\documentclass{article}

\usepackage{microtype}
\usepackage{graphicx}
\usepackage{subfigure}
\usepackage{booktabs} % for professional tables

\usepackage{hyperref}

\usepackage[accepted]{icml2025}

\usepackage{amsmath}
\usepackage{amssymb}
\usepackage{mathtools}
\usepackage{amsthm}

\usepackage[capitalize,noabbrev]{cleveref}

\theoremstyle{plain}
\newtheorem{theorem}{Theorem}[section]
\newtheorem{proposition}[theorem]{Proposition}
\newtheorem{lemma}[theorem]{Lemma}

\theoremstyle{definition}

\newtheorem{assumption}[theorem]{Assumption}
\theoremstyle{remark}
\newtheorem{remark}[theorem]{Remark}

\usepackage[textsize=tiny]{todonotes}

\icmltitlerunning{Randomization Inference and Conformal Selective Borrowing in Hybrid Controlled Trials}

\usepackage{amsmath}
\usepackage{amssymb} % Mathematical formulas
\usepackage{graphicx} % Insert images

\usepackage{xcolor}

\usepackage{booktabs}
\usepackage{threeparttable}
\usepackage{threeparttablex}
\usepackage{array}

\newcommand{\indep}{\perp \!\!\! \perp}
\def\T{{\rm T}}

\def\hg{\hat{\gamma}}
\def\bbP{\mathbb{P}}
\def\bbE{\mathbb{E}}
\def\bbV{\mathbb{V}}
\def\bbI{\mathbb{I}}
\def\bbR{\mathbb{R}}

\def\cS{\mathcal{S}}

\def\cR{\mathcal{R}}
\def\cT{\mathcal{T}}
\def\cC{\mathcal{C}}
\def\cE{\mathcal{E}}
\def\hcE{\hcEE{\gamma}}
\newcommand\hcEE[1]{\hat{\cE}(#1)}

\def\nt{n_1}
\def\nc{n_0}
\def\nr{n_{\cR}}
\def\ne{n_{\cE}}
\def\nre{n}

\def\htr{\hat\tau_{\cR}}
\def\htrc{\hat\tau_{\cR+\cE}}
\def\htrhc{\htau{\gamma}}
\newcommand\htau[1]{\hat\tau_{#1}}

\def\A{\boldsymbol{A}}

\def\cA{\mathcal{A}}

\def\bbPA{\bbP_{\A}}
\def\bbPAs{\bbP_{\A^*}}

\def\hr{\hat{r}}
\def\hp{\hat\pi}

\def\hf{\hat{f}}

\def\sji{s_j^{(i)}}
\def\sjip{s_j^{(i^\prime)}}

\def\full{{\rm full}}
\def\sp{{\rm split}}
\def\jp{{\rm jackknife+}}
\def\cvp{{\rm cv+}}

\def\i{{i^\prime}}
\def\j{{j^\prime}}

\newcommand\hF[1]{{F}_{#1,\nre,M}}
\newcommand\pF[1]{F_{#1,\nre}}
\def\hq{{q}_{\alpha,M}}
\def\pq{q_{\alpha}}

\newcommand\mse[1]{{\rm MSE}(#1)}
\newcommand\hmse[1]{\widehat{\rm MSE}(#1)}

\newcommand\sqb[1]{\{\bbE(\htau{#1}) - \tau\}^2}

\newcommand\var[1]{\bbV(\htau{#1})}

\def\nb{\mathtt{NB}}
\def\fb{\mathtt{FB}}
\def\csb{\mathtt{CSB}}
\def\alsb{\mathtt{ALSB}}

\def\htg{\htau{\gamma}}
\def\hto{\htau{1}}
\def\dg{\delta_{\gamma}}
\def\don{\delta_{1}}
\def\eg{\epsilon_{\gamma}}
\def\eo{\epsilon_{1}}
\def\sg{\sigma_{\gamma}}

\def\pg{\phi_{\gamma}}

\def\kg{\kappa_{\gamma}}

\def\dcap{\Delta}
\def\pcap{\Phi}

\def\la{\lambda}

\def\io{\iota}

\usepackage[ruled,vlined]{algorithm2e}

\begin{document}

\twocolumn[
\icmltitle{Enhancing Statistical Validity and Power in Hybrid Controlled Trials: \\A Randomization Inference Approach with Conformal Selective Borrowing}

% It is OKAY to include author information, even for blind
% submissions: the style file will automatically remove it for you
% unless you've provided the [accepted] option to the icml2025
% package.

% List of affiliations: The first argument should be a (short)
% identifier you will use later to specify author affiliations
% Academic affiliations should list Department, University, City, Region, Country
% Industry affiliations should list Company, City, Region, Country

% You can specify symbols, otherwise they are numbered in order.
% Ideally, you should not use this facility. Affiliations will be numbered
% in order of appearance and this is the preferred way.
\icmlsetsymbol{equal}{*}

\begin{icmlauthorlist}
\icmlauthor{Ke Zhu}{a,b}
\icmlauthor{Shu Yang}{a}
\icmlauthor{Xiaofei Wang}{b}
% \icmlauthor{Firstname4 Lastname4}{sch}
% \icmlauthor{Firstname5 Lastname5}{yyy}
% \icmlauthor{Firstname6 Lastname6}{sch,yyy,comp}
% \icmlauthor{Firstname7 Lastname7}{comp}
% %\icmlauthor{}{sch}
% \icmlauthor{Firstname8 Lastname8}{sch}
% \icmlauthor{Firstname8 Lastname8}{yyy,comp}
% %\icmlauthor{}{sch}
% %\icmlauthor{}{sch}
\end{icmlauthorlist}

\icmlaffiliation{a}{Department of Statistics, North Carolina State University, Raleigh, NC 27695, U.S.A.}
\icmlaffiliation{b}{Department of Biostatistics and Bioinformatics, Duke University, Durham, NC 27710, U.S.A.}
% \icmlaffiliation{sch}{School of ZZZ, Institute of WWW, Location, Country}

\icmlcorrespondingauthor{Shu Yang}{syang24@ncsu.edu}
% \icmlcorrespondingauthor{Firstname2 Lastname2}{first2.last2@www.uk}

% You may provide any keywords that you
% find helpful for describing your paper; these are used to populate
% the "keywords" metadata in the PDF but will not be shown in the document
\icmlkeywords{Machine Learning, ICML}

\vskip 0.3in
]

% this must go after the closing bracket ] following \twocolumn[ ...

% This command actually creates the footnote in the first column
% listing the affiliations and the copyright notice.
% The command takes one argument, which is text to display at the start of the footnote.
% The \icmlEqualContribution command is standard text for equal contribution.
% Remove it (just {}) if you do not need this facility.

\printAffiliationsAndNotice{}  % leave blank if no need to mention equal contribution
%\printAffiliationsAndNotice{\icmlEqualContribution} % otherwise use the standard text.

\begin{abstract}
External controls from historical trials or observational data can augment randomized controlled trials when large-scale randomization is impractical or unethical, such as in drug evaluation for rare diseases. However, non-randomized external controls can introduce biases, and existing Bayesian and frequentist methods may inflate the type I error rate, particularly in small-sample trials where external data borrowing is most critical. To address these challenges, we propose a randomization inference framework that ensures finite-sample exact and model-free type I error rate control, adhering to the ``analyze as you randomize" principle to safeguard against hidden biases. Recognizing that biased external controls reduce the power of randomization tests, we leverage conformal inference to develop an individualized test-then-pool procedure that selectively borrows comparable external controls to improve power. Our approach incorporates selection uncertainty into randomization tests, providing valid post-selection inference. Additionally, we propose an adaptive procedure to optimize the selection threshold by minimizing the mean squared error across a class of estimators encompassing both no-borrowing and full-borrowing approaches. The proposed methods are supported by non-asymptotic theoretical analysis, validated through simulations, and applied to a randomized lung cancer trial that integrates external controls from the National Cancer Database.
\end{abstract}

\section{Introduction}

% RCT, Hybrid control, type I error inflation under small sample size & unmeasured confounding
Randomized controlled trials (RCTs) are the gold standard for making causal inferences on the treatment effect of a new treatment relative to a control treatment. 
However, large RCTs are often infeasible to conduct in practice when the indications of interest involve rare diseases \citep{fda_rare_diseases_2022} or common conditions where few patients are willing to participate due to a lack of equipoise \citep{miller2011equipoise}.
RCTs in such a context often lack sufficient statistical power to detect realistic treatment effect sizes.
Meanwhile, historical studies or large external databases provide real-world data under control conditions, often referred to as external controls (ECs).
By integrating RCT with ECs, hybrid controlled trials have garnered significant interest as an effective approach to enhance the power of RCTs with small sample sizes.
However, most existing methods for hybrid controlled trials rely on model-based or asymptotic $p$-values, which can lead to inflated type I error rates when the randomized sample size is small, or the model is misspecified.
Moreover, since ECs are not randomized, they may systematically differ from randomized controls, even after adjusting for measured confounders. Directly incorporating these ECs may introduce hidden bias, compromising the validity of the statistical inference.
Strictly controlling the type I error rate in hybrid controlled trials, especially with small sample sizes and unmeasured confounding, remains an open problem.

% Randomization Inference: type I error control under small sample size & unmeasured confounding; Power gain under no unmeasured confounding; power loss under unmeasured confounding
To address this problem, we extend the randomization inference framework to hybrid controlled trials. 
To utilize ECs, we use a doubly robust estimator of the average treatment effect (ATE) as the test statistic, which incorporates both RCT and EC data and effectively balances the measured confounders between RCT and EC \citep{li2023improving}.
Then, Fisher randomization tests (FRTs) are performed using only the randomization in the RCT. 
In contrast to the asymptotic inference in \citet{li2023improving}, which relies on (i) large sample sizes for both the RCT and EC, (ii) correct specification of at least one of the two nuisance models, and (iii) no unmeasured confounders, the FRT strictly controls the type I error rate without requiring any of these conditions, thus achieving model-free, finite-sample exact inference.
The validity of the FRT relies solely on the randomization within the RCT, which is typically well-managed by the study design.
Furthermore, we perform a power analysis for FRT in hybrid controlled trials and show that incorporating unbiased ECs with correctly specified models can enhance statistical power. However, EC borrowing is not a free lunch, as including biased ECs may diminish power.

% merit of hybrid controlled trial: selective borrowing to address unmeasured confounding
The power issue motivates us to develop a method that selectively incorporates unbiased ECs rather than indiscriminately borrowing all ECs.
Unlike observational studies, where the assumption of no unmeasured confounders is untestable, a key advantage of hybrid controlled trials is that the bias in ECs can be identified by comparing EC units to randomized control units.
Existing methods mitigate hidden bias by penalized bias estimation and selective borrowing \citep{gao2023integrating}, where selection consistency depends on asymptotic arguments, potentially leading to inferior performance in small samples.

% Conformal Selective Borrowing: (i) individual testing (ii) model-free ditribution-free flexible (iii) finite sample propoerties. 
We propose a novel approach called Conformal Selective Borrowing ($\csb$), which tests the comparability of ECs and selectively incorporates them using conformal inference \citep{vovk2005algorithmic,lei2018distribution}. We measure the bias of each EC using a score function that can flexibly accommodate parametric or machine-learning models. We then calibrate this score to a conformal $p$-value, which test the exchangeability of each EC. These conformal $p$-values are valid in finite samples, distribution-free, and do not depend on the asymptotic properties of models.
$\csb$ offers three advantages: (i) individual borrowing decisions for each EC, (ii) flexibility in using parametric or machine learning models for bias estimation, and (iii) finite-sample guarantees with stable performance in small samples.

In summary, the proposed methods leverage the two key advantages of hybrid controlled trials: (i) randomization within the RCT data allows us to use FRT to control the type I error rate, and (ii) the presence of randomized controls enables us to evaluate bias in ECs using conformal $p$-values, selectively borrow unbiased ECs, and enhance power.
We account for selection uncertainty in FRT and offer valid post-selection inference.
Both FRT and $\csb$ are model-free, distribution-free, and maintain finite-sample exact properties, allowing them to flexibly incorporate state-of-the-art machine learning methods while remaining valid for any sample size or data distribution.
To ensure robust performance across varying bias magnitudes, we propose a data-adaptive procedure for determining the selection threshold to minimize the MSE of the $\csb$ estimator.
Our MSE-guided adaptive threshold offers key advantages: (i) it improves FRT power over RCT-only analysis when EC bias is negligible or detectable; when the bias is non-negligible yet difficult to detect, it may lead to power loss, though FRT still maintains valid Type I error control; (ii) it enables $\csb$ to serve as both a powerful test statistic and an accurate ATE estimator; (iii) the empirical MSE of $\csb$ can be approximated leveraging the RCT-only estimator, making the procedure practically feasible, and we provide a non-asymptotic excess risk bound for its performance.
The advantages of our approach are shown via simulations and a lung cancer RCT with ECs from the National Cancer Database.

\subsection{Related work}
\label{sec:rel1}

% data integration
\textbf{Hybrid controlled trials} aim to integrate ECs to boost RCT efficiency \citep{pocock1976combination}. For an overview of RCT and RWD integration, see \citet{colnet2024causal}. 
A key challenge is biases in ECs, which stem from factors like selection bias, non-concurrency, and measurement error \citep{FDA2023}.
Statistically, biases are categorized as measured and unmeasured confounding. Measured confounding, or covariate shift, refers to systematic differences in observed covariates between RCs and ECs. 
To address measured confounding, covariate balancing techniques such as matching, inverse propensity score weighting, calibration weighting, and their augmented counterparts can be employed \citep{li2023improving,valancius2023causal,li2023efficient}.
When there is unmeasured confounding between RCT and EC, a rich body of literature addresses the hidden bias through various strategies, including test-then-pool \citep{viele2014use,yuan2019design,li2020revisit,ventz2022design,liu2022matching,yang2023elastic,gao2023pretest,dang2022cross}, 
weighted combination \citep{chen2020propensity,chen2021minimax,cheng2021adaptive,li2022conditional,oberst2022understanding,rosenman2023combining,chen2023efficient,karlsson2024robust},
selective borrowing \citep{chen2021combining,li2023frequentist,zhai2022data,gao2023integrating,huang2023simultaneous}, 
bias modeling \citep{stuart2008matching,cheng2023enhancing,li2023confounding,van2024adaptive,yang2024datafusion,gu2024incorporating},
control variates or prognostic adjustment \citep{yang2020combining,guo2022multisource,schuler2022increasing,gagnon2023precise}, 
Bayesian methods \citep{hobbs2011hierarchical,schmidli2014robust,jiang2023elastic,kwiatkowski2024case,alt2024leap,lin2024data,lin2025combining},
and sensitivity analysis \citep{yi2023testing}.
None of them use randomization inference or conformal inference to address unmeasured confounding in hybrid controlled trials with a small sample size.

\textbf{Randomization inference}, introduced by \citet{fisher1935}, provides finite-sample exact $p$-values for any test statistic and is widely endorsed \citep{rosenberger2019randomization, proschan2019re, young2019, bind2020, carter2023regulatory}. 
Randomization tests are useful for small sample trials or complex designs, including cluster experiments with few clusters \citep{rabideau2021randomization} and adaptive experiments \citep{simon2011using, plamadeala2012sequential, nair2023randomization, freidling2024selective}.
Randomization tests have appeared in regulatory guidance documents to ensure type I error rate control in adaptive designs when conventional statistical methods fail \citep{ema2015guideline,fda2019ada,carter2023regulatory}.
For an overview of randomization inference, see \citet{zhang2023randomization} and \citet{ritzwoller2024randomization}. Nevertheless, the randomization inference hasn't been applied to hybrid controlled trials, especially with selective borrowing to address unmeasured confounding.

\textbf{Conformal inference}, or conformal prediction, is a model-free method providing finite-sample valid uncertainty quantification for individual predictions \citep{vovk2005algorithmic}, particularly useful in high-stakes scenarios with black-box machine learning models \citep{angelopoulos2023conformal}.
Two main applications are most relevant to this paper. The first involves using conformal inference to infer individual treatment effects \citep{chernozhukov2021exact, lei2021conformal}. 
The second line is in outlier detection \citep{guan2022prediction, bates2023testing,liang2024integrative}. 
These studies inspire us to treat biased ECs as outliers and use conformal $p$-values to test their exchangeability. 
Our primary goal, however, is to boost FRT power by selectively borrowing unbiased ECs with conformal $p$-values.
The adaptive selection threshold that minimized the estimator's MSE is also a novel approach.

\section{Randomization inference framework}

\subsection{Preliminaries}
\label{sec:pre}

% potential outcome
Consider $\nr$ patients in the RCT, $\ne$ patients in the EC group, and $\nre = \nr+\ne$ patients in total.
Let $S = 1$ for patients in RCT and $S = 0$ for patients in the EC group. 
Let $A$ denote the binary treatment, where $A = 1$ stands for treatment and $A = 0$ stands for control. 
We denote $\cT=\{i:A_i=1,S_i=1\}$, $\cC=\{i:A_i=0,S_i=1\}$, $\cR=\cT\cup\cC$, and $\cE=\{i:S_i=0\}$.
Let $X$ denote the baseline covariates, $Y$ denote the observed outcome, and $Y(0)$ and $Y(1)$ denote the potential outcomes.
In an RCT, we randomize $\nr$ patients into either the treatment or control groups based on the known propensity score $e(x) = \bbP(A=1 \mid X=x, S=1)$. 
This results in $\nt$ patients in the treatment group and $\nc$ patients in the control group.
For $\ne$ patients in the EC group, since all of them are under control, we have $A=0$ for $S=0$.
Let $\pi(x)=\bbP(S=1\mid X=x)$ denote the sampling score of participating in the RCT.
We consider the average treatment effect in the RCT population as our estimand $\tau=\bbE\{Y(1)-Y(0)\mid S=1\}$. 
% RCT-only DR
For RCT data, the following standard identification assumptions are considered \citep{imbens2015causal}.
\begin{assumption}[RCT identification]
\label{ass:rct}
    (i) (Consistency) $Y=AY(1)+(1-A)Y(0)$.
    (ii) (Positivity) $0<e(x)<1$ for all $x$ such that $f_{X|S}(x|1)>0$, where $f_{X|S}(x|s)$ is the conditional p.d.f. of $X$ given $S=s$.
    (iii) (Randomization) $Y(a)\indep A \mid (X, S = 1)$, $a=0,1$.
\end{assumption}
Under Assumption \ref{ass:rct}, $\tau$ is identifiable based on RCT data. 
We denote the conditional outcome mean functions by 
$
\mu_{a}(x)=\bbE(Y\mid X=x,A=a,S=1), a=0,1.
$
We estimate $\mu_{a}(x)$ and $e(x)$ with only RCT data and denote the estimated functions by $\hat{\mu}_{a,\cR}(x)$ and $\hat{e}(x)$, respectively.
An RCT-only doubly robust estimator of $\tau$ is 
\begin{align*}
\htr=\frac{1}{\nr}
\sum_{i=1}^n
&S_i
\bigg[
\hat{\mu}_{1,\cR}(X_i) 
+ \frac{A_i}{\hat{e}(X_i)} \{Y_i-\hat{\mu}_{1,\cR}(X_i)\}
\\
&- \hat{\mu}_{0,\cR}(X_i) 
- \frac{1-A_i}{1-\hat{e}(X_i)} \{Y_i-\hat{\mu}_{0,\cR}(X_i)\}
\bigg],
\end{align*}
which is referred to as the No Borrowing ($\nb$) approach hereafter.
In RCTs, since the propensity score model $e(x)$ is known, $\htr$ is consistent and asymptotically normal regardless of whether $\mu_{a}(x)$ is correctly specified for $a=0,1$. Thus, $\htr$ serves as a model-assisted covariate-adjusted ATE estimator whose asymptotic variance attains the semiparametric efficiency bound if $\mu_{a}(x)$ is correctly specified for $a=0,1$.
% RCT+EC DR
The efficiency of $\htr$ could be further improved by borrowing information from EC data.
To incorporate EC data for estimating $\tau$, many scholars have considered the following assumption \citep{li2023improving}.
\begin{assumption}[Mean exchangeability]
\label{ass:ec}
    $\bbE\{Y(0)\mid X,S=0\}=\bbE\{Y(0)\mid X,S=1\}$.
\end{assumption}
Under Assumptions \ref{ass:rct} and \ref{ass:ec}, $\tau$ could be identified with both RCT and EC data. We estimate $\mu_{0}(x)$ with RCT and EC data and denote the estimated functions by $\hat{\mu}_{0,\cR+\cE}(x)$.
Let $\hp_{\cE}(x)$ denote the estimated sampling score.
The variance ratio between randomized controls and ECs is denoted by $r(x)=\bbV\{Y(0)\mid X=x,A=0,S=1 \}\big/\bbV\{Y(0)\mid X=x,A=0,S=0\}$.
Let $\hr_{\cE}(x)$ denote the estimated variance ratio.
\citet{li2023improving} proposed a doubly robust estimator of $\tau$:
% \begin{align}
% \label{eq:fb}
% \htrc=\frac{1}{\nr}\sum_{i=1}^n
% \bigg[
% &S_i\,\hat{\mu}_{1,\cR}(X_i) 
% + S_i \frac{A_i}{\hat{e}(X_i)} \{Y_i-\hat{\mu}_{1,\cR}(X_i)\}- S_i\,\hat{\mu}_{0,\cR+\cE}(X_i)  \\
% &- \hp_{\cE}(X_i)\frac{S_i(1-A_i)+(1-S_i)\hr_{\cE}(X_i)}{\hp_{\cE}(X_i)\{1-\hat{e}(X_i)\}+ \{1-\hp_{\cE}(X_i)\}\hr_{\cE}(X_i)} \{Y_i-\hat{\mu}_{0,\cR+\cE}(X_i)\}
% \bigg],\nonumber
% \end{align}
\begin{align}
\label{eq:fb}
\htrc&=\frac{1}{\nr}\sum_{i=1}^n
\bigg[
S_i\,\hat{\mu}_{1,\cR}(X_i) 
+ S_i \frac{A_i}{\hat{e}(X_i)} \{Y_i-\hat{\mu}_{1,\cR}(X_i)\}\nonumber\\
&- S_i\,\hat{\mu}_{0,\cR+\cE}(X_i) 
- W_i\{Y_i-\hat{\mu}_{0,\cR+\cE}(X_i)\}
\bigg],
\end{align}
\vspace{-10pt}
$$
W_i = \hp_{\cE}(X_i)\frac{S_i(1-A_i)+(1-S_i)\hr_{\cE}(X_i)}{\hp_{\cE}(X_i)\{1-\hat{e}(X_i)\}+ \{1-\hp_{\cE}(X_i)\}\hr_{\cE}(X_i)}.
$$
$\htrc$ is referred to as the Full Borrowing ($\fb$) approach hereafter.
The term ``Full" here refers to incorporating the full set of ECs to construct $\htrc$, while down-weighting those ECs based on similarity measured by $X$, thereby addressing bias caused by observed confounders.
$\htrc$ is consistent and asymptotically normal if either (i) $\mu_{a}(x)$ is correctly specified for $a=0,1$, or (ii) both $\pi(x)$ and $e(x)$ are correctly specified. If all models for $\mu_{a}(x)$, $a=0,1$, $\pi(x)$, and $e(x)$ are correctly specified, $\htr$ achieves the semiparametric efficiency bound. 

% motivation: small sample & unmeasured confounding
However, asymptotic inference for $\htrc$ may be invalid due to three main reasons: (i) it assumes $\nr \rightarrow \infty$, which contradicts the motivation for EC borrowing, where the sample size of the RCT is typically small; (ii) it relies on the correct specification of at least one of the two nuisance models, which may be violated because sophisticated models are difficult to work with under small sample sizes; and (iii) it depends on Assumption \ref{ass:ec}, which may be violated due to unmeasured confounders.
To address these issues, we consider a finite-sample exact randomization inference framework that maintains strict type I error rate control even if all models are misspecified and Assumption \ref{ass:ec} fails.
We consider $\htr$ and $\htrc$ as candidate test statistics and propose a new class of test statistics in Section \ref{sec:csb} to achieve improved power across various scenarios.

\subsection{Fisher randomization test}

In the randomization inference framework, we are conditional on the potential outcomes $Y_i(a)$ and covariates $X_i$ for $i \in \cR \cup \cE$, and consider the randomized assignment $\A = (A_1, \ldots, A_{\nre})$ as the sole source of randomness.
Since $A_i$ for $i \in \cR$ is well controlled and known in the RCT, we can leverage this advantage to guarantee the validity of inference without any additional assumptions.
Let $\cA$ denote the set of all possible assignments generated by the actual RCT design.
Since all external units are under control, we have $A_i=0$ for $i\in\cE$.
Randomization inference accommodates not only Bernoulli trials with $A_i \overset{i.i.d.}\sim \text{Bernoulli}(p)$ for $i \in \cR$ but also complex designs like covariate-adaptive randomization \citep{rosenberger2015randomization}.

Consider Fisher's sharp null hypothesis $H_0: Y_i(0)=Y_i(1)$, $\forall i\in\cR$, which states no treatment effect for any units in RCT. Based on $H_0$, we could impute all potential outcomes $Y_i^{\text{imp}}(0)=Y_i^{\text{imp}}(1)=Y_i$ for $i\in\cR$.
Let $T(\A)$ denote the test statistic, which depends on the assignment $\A \in \cA$. 
$T(\A)$ could be $|\htr(\A)|$, $|\htrc(\A)|$, or the estimator introduced in Section \ref{sec:csb}.
The theoretical guarantee of type I error rate control holds for \textit{any test statistic}, including those involving ECs, even if these ECs have hidden biases. This is one of the key merits of randomization inference.
We define the $p$-value for measuring the extremeness of the observed $T(\A)$ against $H_0$ as 
$$
p^{\rm FRT}=\bbPAs\left\{T(\A^*)\geq T(\A)\right\},
$$
where $\A^*\in\cA$ has the same distribution as $\A$ and is independent of $\A$, and $\bbPAs$ is taken over the distribution of $\A^*$.

\begin{theorem}
\label{thm:frt}
Under $H_0$, for $\alpha\in(0,1)$, we have $\bbPA(p^{\rm FRT}\leq\alpha)\leq\alpha$,
where $\bbPA$ is taken over the distribution of $\A$.
If we further assume that $T(\A)$ takes distinct values for different $\A\in\cA$, then we have
$
\bbPA(p^{\rm FRT}\leq\alpha)
=\lfloor\alpha|\cA|\rfloor/|\cA|
>\alpha - 1/|\cA|,
$
where $\lfloor x \rfloor$ represents the greatest integer less than or equal to $x$.
\end{theorem}

In practice, we use Monte Carlo to approximate $p^{\rm FRT}$.
Based on the RCT's actual randomization, we generate the new assignment $A_i^b$ for $i \in \cR$ and set $A_i^b \equiv 0$ for $i \in \cE$ since the randomization in the RCT does not affect the assignments of the ECs. 
A caveat is that the assignment of ECs should not be permuted, as this would violate the ``analyze as you randomize" principle and compromise the validity of the FRT.
The new assignment vector is denoted as $\A^{b} = (A_1^b, \ldots, A_{\nre}^b)$.
We generate assignments for $B$ times and obtain
$\hat{p}^{\rm FRT}=\big[\sum_{b=1}^B \bbI\{T(\A^b)\geq T(\A)\}+1\big] / (B+1),$
where the ``$+1$" term accounts for $\A$ itself.

Theorem \ref{thm:frt} shows that FRT exactly controls the type I error rate in finite samples, regardless of Assumption \ref{ass:ec}, because, under $H_0$, the reference distribution is derived from true randomization, which is well-controlled in clinical trials. However, the power of FRT heavily depends on the choice of test statistic, making it the most critical decision in randomization inference.

\subsection{Model-based power analysis}

There are two primary approaches for conducting a power analysis of FRT: model-based or simulation-based \citep{rosenberger2015randomization}.
We first perform a model-based power analysis under Assumption \ref{ass:ec}, highlighting how low variance of a consistent test statistic enhances the power of FRT. In the following section, we conduct a simulation-based power analysis for a more challenging scenario where Assumption \ref{ass:ec} does not hold, showing that the bias of an inconsistent test statistic reduces the power of FRT.

Let $M$ denote the total number of possible assignments, $\hF{1}(t)=\bbPA(T(\A)\leq t)$ denote the randomization distribution of $T(\A)$, and $\hF{0}(t)=\bbPAs(T(\A^*)\leq t)$ denote the reference distribution of $T(\A^*)$ under $H_0$. Both $\hF{1}$ and $\hF{0}$ are discrete in finite samples.
To apply empirical process theory and derive asymptotic rates for testing power, we assume continuous super-population distributions $\pF{1}$ and $\pF{0}$, with $\hF{1}$ and $\hF{0}$ representing the empirical distribution functions based on $M$ independent samples drawn from $\pF{1}$ and $\pF{0}$, respectively.
In cases where these assumptions do not hold, FRT still controls the type I error rate, and we will investigate its power through simulation in Section \ref{sec:sim}.
Based on those notations, the $p$-value and the power can be expressed as
$
p^{\rm FRT}=\bbPAs\left\{T(\A^*)\geq T(\A)\right\}=1-\hF{0}\big(T(\A)\big),
$
and $\psi_{\nre,M} = \bbPA(p^{\rm FRT}\leq\alpha)
=\bbPA\{1-\hF{0}\big(T(\A)\big)\leq\alpha\}
=1-\hF{1}\big(\hF{0}^{-1}(1-\alpha)\big)$.

\begin{theorem}
\label{thm:power}
For fixed $\nre>0$, suppose

(a) There are continuous cumulative distribution functions (c.d.f.) $\pF{0}$ and $\pF{1}$, such that $\hF{0}$ and $\hF{1}$ are the empirical distribution functions based on $M$ independent samples drawn from $\pF{0}$ and $\pF{1}$, respectively.

(b) There is $\sigma_\nre>0$ and a continuous c.d.f. $F$ such that $\pF{0}(t)=F(t / \sigma_\nre)$ for all $t \in \bbR$.

(c) For ATE $\tau$, $\pF{1}(t)=\pF{0}(t-\tau)=F\big((t-\tau) / \sigma_\nre\big)$, for all $t \in \mathbb{R}$.

For $0 < \iota < 0.5$ and sufficiently large $M$, 
$$
\bbE(\psi_{\nre,M}) \geq 1 - F(F^{-1}(1-\alpha) - \tau / \sigma_\nre) - O(M^{-0.5+\iota}),
$$
where $\bbE$ is over $M$ independent samples from $\pF{0}$ and $\pF{1}$.
\end{theorem}

For a given $\tau \neq 0$, significance level $\alpha$, and design with possible assignments $M$, Theorem \ref{thm:power} shows that the power of the FRT also depends on the variance of the test statistic, $\sigma_\nre$. Under Assumptions \ref{ass:rct} and \ref{ass:ec}, and with all working models correctly specified, $\htrc$ is consistent and has a variance that is less than or equal to that of $\htr$ \citep{li2023improving}.
Thus, when there is no hidden bias, using $\htrc$ as the test statistic improves the power of the FRT compared to $\htr$, as shown in subplot (B) of Figure \ref{fig:sim_const_ne50_pdist}.

\subsection{Simulation-based power analysis}

\begin{figure*}[ht]
    \centering
    \includegraphics[width=1\linewidth]{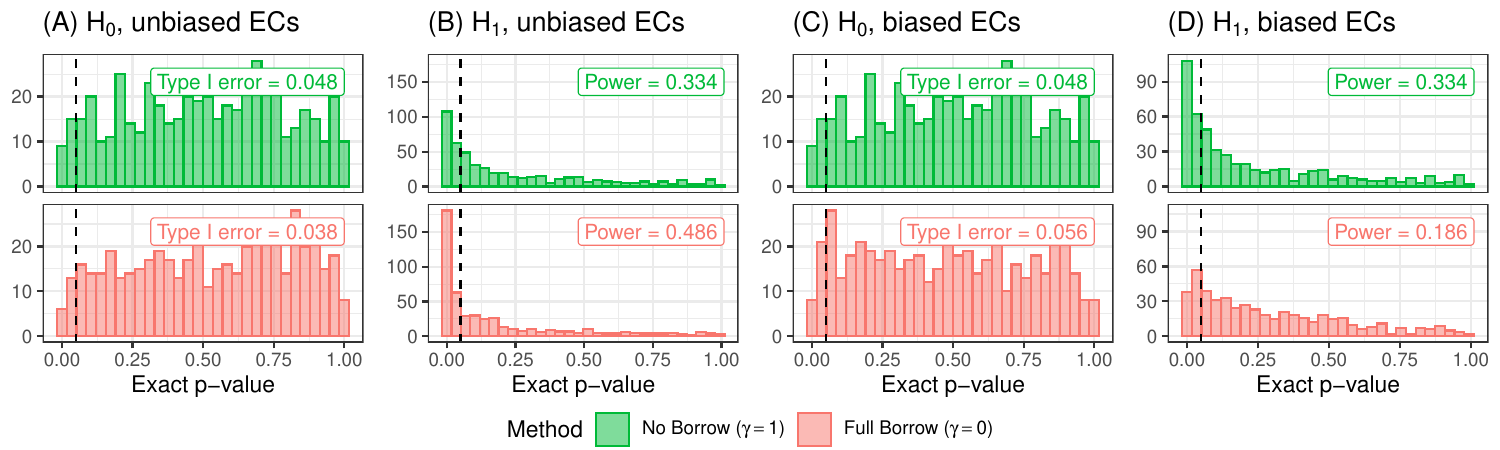}
    \caption{{Simulated distributions of $p$-values under $H_0$ and $H_1$.}}
    \label{fig:sim_const_ne50_pdist}
\end{figure*}

When unmeasured confounding exists between RCT and EC data, Assumption \ref{ass:ec} is violated, rendering $\htrc$ inconsistent.
In such cases, asymptotic inference based on $\htrc$ is invalid and fails to control the type I error rate.
In contrast, since Theorem \ref{thm:frt} holds for any test statistic, FRT can still control the type I error with the inconsistent test statistic $\htrc$, highlighting a core merit of FRTs.
However, the violation of Assumption \ref{ass:ec} subsequently causes Assumption (c) in Theorem \ref{thm:power} to be unfulfilled, rendering FRT with $\htrc$ unable to achieve a power improvement over FRT with $\htr$. Furthermore, employing $\htrc$ as the test statistic results in a substantial loss of power compared to using $\htr$, as illustrated in subplot (D) of Figure \ref{fig:sim_const_ne50_pdist}.

%Based on the power analysis, we find that, compared to $\htr$, $\htrc$ improves the power of the FRT by leveraging all ECs under Assumption \ref{ass:ec}, but results in a severe decrease in power when Assumption \ref{ass:ec} does not hold.
The trade-off between $\htr$ and $\htrc$ generally arises between a causal estimator that ignores additional information and assumptions and one that incorporates them but risks bias if the assumptions fail \citep{rothenhausler2020model,rothenhausler2021anchor}.
In the next section, instead of choosing between $\htr$ and $\htrc$, we construct a class of ATE estimators, $\htrhc$, indexed by a tuning parameter $\gamma$ and encompassing $\htr$ and $\htrc$ as special cases. We then propose a data-adaptive procedure to select $\gamma$ that minimizes the MSE of $\htrhc$, thereby enhancing the power of FRT by using $\htrhc$ as the test statistic.

\section{Conformal Selective Borrowing}
\label{sec:csb}

\subsection{A class of estimators}

Motivated by heterogeneous scenarios where some ECs satisfy Assumption \ref{ass:ec} while others do not, we propose an individualized test-then-pool approach that leverages conformal inference to select comparable ECs.
The conformal $p$-value $p_{j}^* \in (0,1]$ is used to test the exchangeability of each EC $j \in \cE$. The selected EC set is then defined as $\hcE = \{j \in \cE : p_{j}^* > \gamma\}$, where $\gamma\in[0,1]$ is a selection threshold. Substituting $\cE$ with $\hcE$ in \eqref{eq:fb}, we obtain the Conformal Selective Borrowing ($\csb$) estimator:
% \begin{align}
% \label{eq:sel}
% \htrhc=&\frac{1}{\nr}\sum_{i=1}^n
% \bigg[
% S_i\,\hat{\mu}_{1,\cR}(X_i) 
% + S_i \frac{A_i}{\hat{e}(X_i)} \{Y_i-\hat{\mu}_{1,\cR}(X_i)\} - S_i\,\hat{\mu}_{0,\cR+\hcE}(X_i) \\
% &- \hat\pi_{\hcE}(X_i)\frac{S_i(1-A_i)+(1-S_i)\bbI\{i\in\hcE\}\hat{r}_{\hcE}(X_i)}{\hat\pi_{\hcE}(X_i)\{1-\hat{e}(X_i)\}+ \{1-\hat\pi_{\hcE}(X_i)\}\hat{r}_{\hcE}(X_i)} \{Y_i-\hat{\mu}_{0,\cR+\hcE}(X_i)\}
% \bigg]. \nonumber
% \end{align}
\begin{align}
\label{eq:sel}
&\htrhc=\frac{1}{\nr}\sum_{i=1}^n
\bigg[
S_i\,\hat{\mu}_{1,\cR}(X_i) 
+ S_i \frac{A_i}{\hat{e}(X_i)} \{Y_i-\hat{\mu}_{1,\cR}(X_i)\} \nonumber\\
&- S_i\,\hat{\mu}_{0,\cR+\hcE}(X_i) - V_i \{Y_i-\hat{\mu}_{0,\cR+\hcE}(X_i)\}
\bigg],
\end{align}
%\vspace{-10pt}
% \begin{align*}
% &V_i = \hat\pi_{\hcE}(X_i)\\
% &\times\frac{S_i(1-A_i)+(1-S_i)\bbI\{i\in\hcE\}\hat{r}_{\hcE}(X_i)}{\hat\pi_{\hcE}(X_i)\{1-\hat{e}(X_i)\}+ \{1-\hat\pi_{\hcE}(X_i)\}\hat{r}_{\hcE}(X_i)}.
% \end{align*}
\begin{align*}
&V_i = \hat\pi_{\hcE}(X_i)\\
&\times\frac{S_i(1-A_i)+(1-S_i)\bbI\{i\in\hcE\}\hat{r}_{\hcE}(X_i)}{\hat\pi_{\hcE}(X_i)\{1-\hat{e}(X_i)\}+ \{1-\hat\pi_{\hcE}(X_i)\}\hat{r}_{\hcE}(X_i)}.
\end{align*}
$\csb$ represents a class of ATE estimators: when $\gamma=1$, no ECs borrowed and $\hcEE{1} = \varnothing$, we have $\htau{1} \equiv \htr$; when $\gamma=0$, all ECs borrowed and $\hcEE{0} = \cE$, we have $\htau{0} = \htrc$. 
%Hereafter, we will use $\htau{1}$ and $\htau{0}$ to refer to the $\nb$ and $\fb$ ATE estimators, respectively.
For $0 < \gamma < 1$, $\htau{\gamma}$ balances the trade-off between borrowing more ECs with a smaller $\gamma$ and discarding more ECs with a larger $\gamma$.
By using $T(\A) = |\htrhc|$ as the test statistic for FRT and allowing $\hcE$ to vary with resampling $\A$ in FRT could account for selection uncertainty and provide valid post-selection inference.
%Figure \ref{fig:sim_const_ne50_pdist} shows that the $\csb$ can improve power compared to the $\nb$, regardless of whether Assumption \ref{ass:ec} holds for all ECs.
The following sections introduce various conformal $p$-values and a data-adaptive procedure for selecting $\gamma$ to minimize the MSE of $\htau{\gamma}$.

\subsection{Conformal \textit{p}-value}

\textbf{Split conformal \textit{p}-value.} 
%To compute full conformal $p$-values for all ECs $j\in\cE$, the prediction model must be refit $\ne$ times, which is time-consuming for large EC samples. 
%Thus, we consider split conformal inference \citep{papadopoulos2002inductive}, requiring only one model fit while preserving validity.
{We first consider split conformal inference \citep{papadopoulos2002inductive}.}
We randomly split $\cC$ into a calibration set $\cC_1$ and a training set $\cC \setminus\cC_1$ according to a prespecified sample size ratio, for example, $1:3$. 
%We can still use the absolute residual as the score function:
{We use a score function $s(x,y)$ to measure the ``nonconformity" of $(x,y)$. 
For example, we can use the absolute residual as the score function:}
$s_i=|Y_i-\hf_{-\cC_1}(X_i)|$ for $i\in\cC_1$ and $s_j=|Y_j-\hf_{-\cC_1}(X_j)|$,
% \begin{gather*}
% s_i=|Y_i-\hf_{-\cC_1}(X_i)|,\quad i\in\cC_1,\\
% s_j=|Y_j-\hf_{-\cC_1}(X_j)|,
% \end{gather*}
where $\hf_{-\cC_1}(x)$ is a prediction model fitted by the training set $\cC \setminus\cC_1$.
Intuitively, if $(X_j,Y_j)$ is not exchangeable (see Remark \ref{rm:ex} for a formal definition) with $\{(X_i,Y_i)\}_{i\in\cC_1}$, $s_j$ should be large compared to $\{s_i\}_{i\in\cC_1}$. 
Thus, we define the split conformal $p$-value as the proportion of $\{s_i\}_{i\in\cC_1}$ that are larger than $s_j$, that is, $p_{j}^\sp= \{\sum_{i\in\cC_1}\bbI(s_i\geq s_j)+1\}/(|\cC_1|+1)$, where $\bbI$ is the indicator function, and the ``+1" accounts for including $s_j$ itself.
If $p_{j}^{\sp}$ is smaller than a threshold $\gamma$, we reject the hypothesis of exchangeability and discard EC $j$. The following theoretical guarantee states that if EC $j$ is exchangeable, the rejection rate is less than $\gamma$.
% $$
% p_{j}^\sp= \frac{\sum_{i\in\cC_1}\bbI(s_i\geq s_j)+1}{|\cC_1|+1}.
% $$

\begin{proposition}
\label{prop:scf}
For $j\in\cE$, suppose that $(X_j,Y_j)$ and $\{(X_i,Y_i)\}_{i\in\cC}$ are exchangeable. For $\gamma\in(0,1)$, we have 
$
\bbP(p_{j}^\sp\leq \gamma)\leq\gamma.
$
If $s_j$ and $\{s_i\}_{i\in\cC}$ have distinct values, we have
$
\bbP(p_{j}^\sp\leq \gamma)=\big\{\lfloor\gamma(|\cC_1|+1)\rfloor\big\}\,\big/\,(|\cC_1|+1)>\gamma - 1/(|\cC_1|+1).
$
% $$
% \bbP(p_{j}^\sp\leq \gamma)=\frac{\lfloor\gamma(|\cC_1|+1)\rfloor}{|\cC_1|+1}>\gamma - \frac{1}{|\cC_1|+1}.
% $$
\end{proposition}

\begin{remark}[Definition of exchangeability]
\label{rm:ex}
The random variables $z_1,\ldots,z_n$ are exchangeable if, for any permutation $\omega$ of $1,\ldots,n$, the random variables $z_{\omega(1)},\ldots,z_{\omega(n)}$ have the same joint distribution as $z_1,\ldots,z_n$.
The i.i.d. assumption is stronger than exchangeability, as the latter can hold with dependence \citep{shafer2008tutorial}.
The exchangeability required by conformal inference is stronger than the mean exchangeability (Assumption \ref{ass:ec}), which allows the construction of a statistically valid estimator within the asymptotic inference framework.
\end{remark}

\textbf{CV+ \textit{p}-value.} 
While split conformal $p$-values are computationally efficient, they lose statistical efficiency due to data splitting.
CV+ \citep{barber2021predictive} fully utilize training data and remain computationally feasible.
We randomly split $\cC$ into $K$ disjoint folds: $\cC=\cup_{k=1}^K\cC_k$.
We use the training set $\cC \setminus\cC_k$ to fit prediction models $\hf_{-\cC_k}(x)$ and use the absolute residual as the score function:
$s_i=|Y_i-\hf_{-\cC_{k(i)}}(X_i)|$ and $\sji=|Y_j-\hf_{-\cC_{k(i)}}(X_j)|$ for $i\in\cC$,
% \begin{gather*}
% s_i=|Y_i-\hf_{-\cC_{k(i)}}(X_i)|,\quad i\in\cC,\\
% \sji=|Y_j-\hf_{-\cC_{k(i)}}(X_j)|, \quad i\in\cC,
% \end{gather*}
where $k(i)\in\{1,\ldots,K\}$ is a function that indicates $i\in\cC_k$.
Thus, for $i\neq i^\prime$ and $k(i)=k(i^\prime)$, we have $\sji=\sjip$.
We define the CV+ $p$-value as the proportion of $\{s_i\}_{i\in\cC}$ that are larger than the corresponding $\{\sji\}_{i\in\cC}$, that is, $p_{j}^\cvp= \{\sum_{i\in\cC}\bbI(s_i\geq \sji)+1\}/(|\cC|+1)$.
%The following proposition is a generalization of Proposition \ref{prop:jp}, where $K=|\cC|$.

\begin{proposition}
\label{prop:cvp}
For $j\in\cE$, suppose that $(X_j,Y_j)$ and $\{(X_i,Y_i)\}_{i\in\cC}$ are exchangeable. For $\gamma\in(0,1)$, we have 
$
\bbP(p_{j}^\cvp\leq\gamma)\leq
2\gamma+\big\{(1-2\gamma)(m-1) -1\big\}\,\big/\,(|\cC|+m)<
2\gamma +\big(1-K/|\cC|\big)\,/\,(K+1),
$
where $m = |\cC| / K$ is assumed to be an integer for simplicity.
\end{proposition}

\subsection{Adaptive selection threshold}
\label{sec:ada_t}

Since we construct $p_{j}^*$ individually and make borrowing decisions collectively, one might consider choosing a selection threshold $\gamma$ that controls the family-wise type I error rate or false discovery rate for testing the exchangeability of all ECs \citep{bates2023testing}. However, in our context, the power of the conformal tests is of greater concern. The classical test-then-pool approach has been criticized for its low power in detecting hidden bias, especially with small randomized control sample sizes \citep{li2020revisit}.
Even with effective control of the family-wise type I error rate, low-power conformal tests can allow many biased ECs to be incorrectly borrowed, increasing the MSE of $\htrhc$ and reducing the power of the FRT.
Therefore, we propose a data-adaptive procedure to directly minimize the MSE of $\htrhc$.

We decompose $\mse{\gamma} \equiv \bbE(\htau{\gamma} - \tau)^2 = \sqb{\gamma} + \var{\gamma}$. The main challenge lies in estimating the squared bias $\{\bbE(\htau{\gamma} - \tau)\}^2$ as the true $\tau$ is unknown. Fortunately, since the $\nb$ estimator $\htau{1}$ is consistent for $\tau$, we approximate $\{\bbE(\htau{\gamma} - \tau)\}^2$ by $\{\bbE(\htau{\gamma} - \htau{1})\}^2 = \bbE(\htau{\gamma} - \htau{1})^2 - \bbV(\htau{\gamma} - \htau{1})$. We then use $(\htau{\gamma} - \htau{1})^2$ to estimate $\bbE(\htau{\gamma} - \htau{1})^2$ and apply bootstrap to estimate $\bbV(\htau{\gamma})$ and $\bbV(\htau{\gamma} - \htau{1})$. Combining these provides the estimated MSE for each $\gamma$ over finite grids, and we select the $\gamma$ that minimizes it. The complete procedure is detailed in Algorithm \ref{alg:gamma}.

\begin{algorithm}[ht]
\caption{Adaptive Selection Threshold}
\label{alg:gamma}
\KwIn{Grid $\Gamma=\{0, 0.1, \dots, 1\}$; bootstrap times $L$.}

%\textbf{Step 1} (ATE Estimation for Original and Bootstrap Samples)

\For{$\gamma \in \Gamma$ }{
Compute $\htau{\gamma}$ from the original sample. 
    
    \For{$l = 1, \ldots, L$}{
    Compute $\htau{\gamma}^{(l)}$ from the $l$-th bootstrap sample.
    }
}

%\textbf{Step 2} (MSE Calculation for Each $\gamma$)

% \For{$\gamma \in \Gamma\setminus\{1\}$} 
\For{$\gamma \in \Gamma$} 
{

% $\widehat{\bbV}(\htau{\gamma} - \htau{1})=(L-1)^{-1} \sum_{l=1}^L \left\{ (\htau{\gamma}^{(l)} - \htau{1}^{(l)}) - L^{-1} \sum_{l^\prime=1}^L (\htau{\gamma}^{(l^\prime)} - \htau{1}^{(l^\prime)}) \right\}^2$.

Compute $\widehat{\bbV}(\htau{\gamma} - \htau{1})$ using $\htau{\gamma}^{(l)} - \htau{1}^{(l)}$.
    
% $\widehat{\bbV}(\htau{\gamma}) = {(L-1)}^{-1} \sum_{l=1}^L \left( \htau{\gamma}^{(l)} - L^{-1} \sum_{l^\prime=1}^L \htau{\gamma}^{(l^\prime)} \right)^2$.

Compute $\widehat{\bbV}(\htau{\gamma})$ using $\htau{\gamma}^{(l)}$.
    
$\hmse{\gamma} = (\htau{\gamma} - \htau{1})^2 - \widehat{\bbV}(\htau{\gamma} - \htau{1}) + \widehat{\bbV}(\htau{\gamma})$.
}

% $\hmse{1} = {(L-1)}^{-1} \sum_{l=1}^L \left( \htau{1}^{(l)} - L^{-1} \sum_{l^\prime=1}^L \htau{1}^{(l^\prime)} \right)^2$.

%\textbf{Step 3} (Optimal Threshold Selection)

\KwOut{$\hg=\arg\min_{\gamma \in \Gamma} \hmse{\gamma}$}

\end{algorithm}

We theoretically analyze the procedure from a non-asymptotic perspective \citep{wainwright2019high}.
Decomposing $\htg = \tau + \dg+ \eg$, where $\dg\equiv\bbE(\htg)-\tau$ and $\bbE(\eg)=0$. Let $\kg^2\equiv\bbV(\htg-\hto)=\bbV(\eg-\eo)$ and $\sg^2\equiv\bbV(\htg)=\bbV(\eg)$.

% \newpage
\begin{theorem}
\label{thm:mse2}
For fixed $\nre > 0$ and $\gamma \in \Gamma$, let $\eg$ be a centered sub-Gaussian variable with parameter $\pg>0$, i.e., $\bbE{\exp(\la \eg)} \leq \exp(\pg^2\la^2/2)$ for all $\la \in \bbR$. For $\io > 0$, there exists $c > 0$ such that with probability at least $1 - 4\io$:
\begin{align*}
&\max_{\gamma\in\Gamma}\left|
\hmse{\gamma} -
\mse{\gamma}
\right|
\leq\; 
c\dcap|\don|+
c\dcap\pcap\sqrt{\log\left(|\Gamma|/\iota\right)}\\
&\quad\quad+\max\left\{c\pcap^2\sqrt{\log\left(|\Gamma|/\iota\right)},\;c\pcap^2\log\left(|\Gamma|/\iota\right)\right\}\\
&\quad\quad+\max_{\gamma\in\Gamma} |\widehat{\bbV}(\htg - \hto)-\kg^2|
+ \max_{\gamma\in\Gamma} |\widehat{\bbV}(\htg)-\sg^2|,
\end{align*}
where $\dcap=\max_{\gamma\in\Gamma}|\dg|$, $\pcap=\max_{\gamma\in\Gamma}\pg$, $\don$ is the bias of $\htau{1}$, and $|\Gamma|$ is the cardinality of $\Gamma$.
\end{theorem}

Theorem \ref{thm:mse2} shows that the discrepancy between the estimated and true MSE vanishes if $\dcap$ is bounded and the bias of the consistent estimator $\htau{1}$, the maximum standard deviation proxy $\pcap$, and the variance estimation errors are sufficiently small.

\begin{theorem}
\label{cor:mse3}
Under the same assumptions as in Theorem \ref{thm:mse2}, for any $\io > 0$, there exists a constant $c > 0$ such that, with probability at least $1 - 8\io$, the following holds:
% \begin{align*}
% (\htau{\hg} - \tau)^2
% -\min_{\gamma\in\Gamma} (\htau{\gamma} - \tau)^2 
% \leq\; 
% &2c\dcap|\don|+
% 2c\dcap\pcap\sqrt{\log\left(|\Gamma|/\iota\right)}
% +2\max\left\{c\pcap^2\sqrt{\log\left(|\Gamma|/\iota\right)},\;c\pcap^2\log\left(|\Gamma|/\iota\right)\right\}\\
% &+2\max_{\gamma\in\Gamma} |\widehat{\bbV}(\htg - \hto)-\kg^2|
% + 2\max_{\gamma\in\Gamma} |\widehat{\bbV}(\htg)-\sg^2|.
% \end{align*}
\begin{align*}
&(\htau{\hg} - \tau)^2
-\min_{\gamma\in\Gamma} (\htau{\gamma} - \tau)^2 
\leq\; 
2c\dcap|\don|+
2c\dcap\pcap\sqrt{\log\left(|\Gamma|/\iota\right)}
\\ &\quad\quad+2\max\left\{c\pcap^2\sqrt{\log\left(|\Gamma|/\iota\right)},\;c\pcap^2\log\left(|\Gamma|/\iota\right)\right\}\\
 &\quad\quad+2\max_{\gamma\in\Gamma} |\widehat{\bbV}(\htg - \hto)-\kg^2|
+ 2\max_{\gamma\in\Gamma} |\widehat{\bbV}(\htg)-\sg^2|.
\end{align*}
\end{theorem}

Theorem \ref{cor:mse3} provides a bound for the excess risk of $\htau{\hg}$ in comparison to the oracle estimator.
Although $\htau{\hg}$ generally outperforms $\htau{1}$ in terms of MSE, it may exhibit excess risk in certain challenging cases, as shown in Figure \ref{fig:sim_const_ne50_main} (C) in the simulation. 
This phenomenon highlights that $\htau{\hg}$ behaves similarly to the Hodges estimator \citep{LeCam1953} and to integrated estimators in data fusion \citep{yang2023elastic,oberst2022understanding}: improving upon the baseline estimator (here, the No Borrow estimator) in certain regions of the parameter space (where there is no bias in ECs) inevitably leads to worse performance in other regions (where the bias in ECs is difficult to detect).
FRT still controls the type I error rate even if excess risk is present or the assumptions in Theorem \ref{thm:mse2} are not satisfied.

\section{Simulation}
\label{sec:sim}

\begin{figure*}[ht]
    \centering
    \includegraphics[width=1\linewidth]{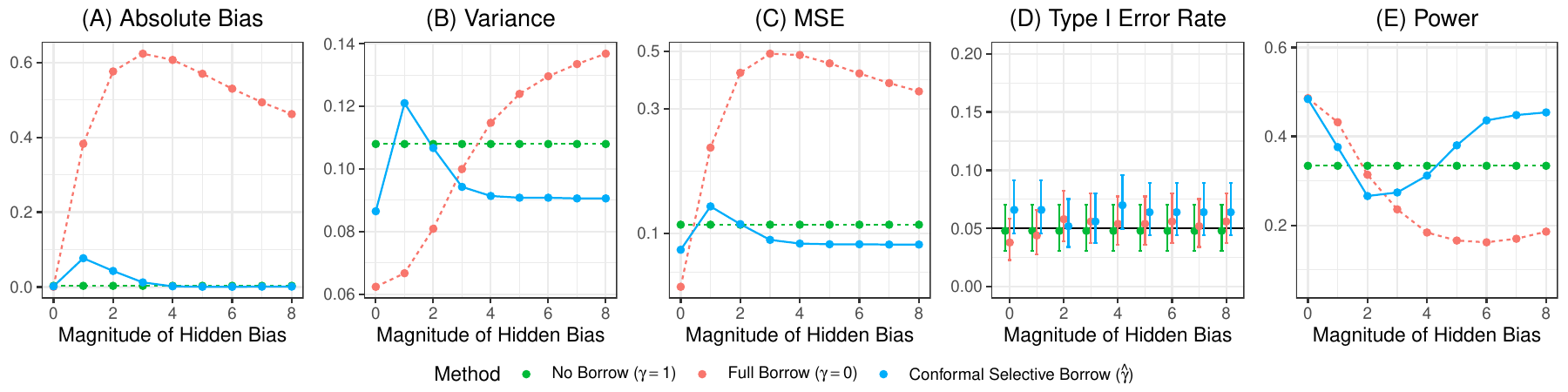}
    \includegraphics[width=1\linewidth]{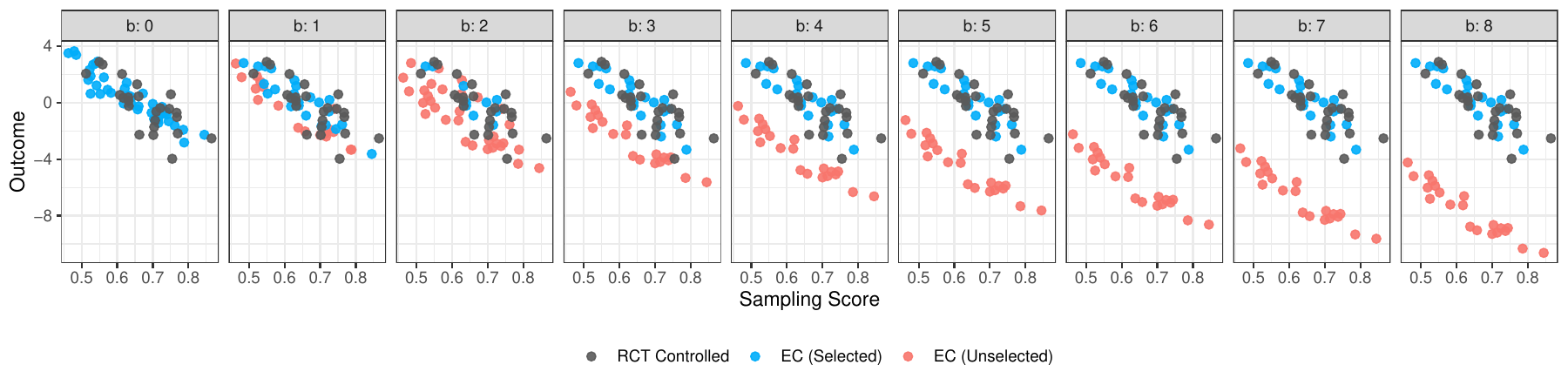}
    \caption{{Simulation results across different hidden bias magnitudes $b$.}}
    \label{fig:sim_const_ne50_main}
\end{figure*}

% setup
We conduct simulations to evaluate the repeated sampling performance of the proposed methods under small sample sizes and varying magnitudes of hidden bias, including challenging cases where separating biased ECs is difficult. 
Specifically, the sample sizes for the randomized treatment, randomized control, and EC groups are set as $(\nt, \nc, \ne) = (50, 25, 50)$.
Similar results for a larger EC sample size ($\ne = 300$) are included in the Appendix.
%We generate covariates $X \sim N(0, I_p)$ with dimension $p = 2$ and $I_p$ is the identity matrix. 
We generate covariates $X \sim \text{Unif}(-2,2)$ with dimension $p = 2$. 
The sampling indicator $S \sim \text{Bernoulli}(\pi(X))$ is generated with $\pi(X) = \{1 + \exp{(\eta_0 + X^{\T}\eta)}\}^{-1}$, where $\eta_0$ is chosen to ensure $\mathbb{E}(S) = \nr/\nre$, and $\eta = (0.1, 0.1)$. The assignment is generated by $A \sim \text{Bernoulli}(\nt/\nr)$ for $S = 1$ and $A = 0$ for $S = 0$.
Let $\varepsilon\sim N(0,1)$ denote the noise.
For the RCT sample ($S = 1$), we generate the potential outcomes as $Y(0) = X^\T\beta_0+\varepsilon$ with $\beta_0 = (1, 1)$, and $Y(1) = \tau_0+X^\T\beta_1+\varepsilon$ with $\tau_0=0.4$ and $\beta_1 = (2, 2)$. 
For the EC sample ($S = 0$), we consider two scenarios: (i) the scenario without hidden bias, where $Y(0)=X^\T\beta_0+0.5\varepsilon$; (ii) the scenario where part of the ECs have hidden bias $b$, where a random proportion $\rho$ of the ECs is biased, with $Y(0) = - b+X^\T\beta_0 +0.5\varepsilon$, and the remaining proportion $(1-\rho)$ are unbiased, with $Y(0)=X^\T\beta_0+0.5\varepsilon$.
We consider proportions of biased ECs $\rho = 50\%$ and magnitudes of hidden bias $b = 1, 2, \ldots,8$.
Note that hidden bias refers to bias that remains due to unmeasured confounders, even after balancing the observed covariates.
Under the alternative hypothesis, the observed outcome is $Y = AY(1) + (1 - A)Y(0)$; under the sharp null hypothesis, the observed outcome is $Y = Y(0)$.
% method
We consider $\nb$, $\fb$, and $\csb$ with the adaptive selection threshold as estimators of $\tau$ and test statistics for FRT. 
We also consider Adaptive Lasso Selective Borrowing ($\alsb$) by \citet{gao2023integrating}. Given its higher computational cost (approximately 10 times slower than $\csb$), we omit FRTs for this method and instead compare $\csb$+FRT with $\alsb$+asymptotic inference in the Appendix.
CV+ $p$-values are used with 10 folds. 
%We use Algorithm \ref{alg:gamma} to determine $\hg$. We also consider various fixed thresholds $\gamma=0.4, 0.6, 0.8$ in the Appendix. 
We set $B=5000$ to approximate $p^{\rm FRT}$ and replicate the simulation 500 times per scenario.

% results
% no bias: bias, mse, type I, power
% bias: bias, mse, type I, power
Figure \ref{fig:sim_const_ne50_main} displays performance metrics for $b=0,1,\ldots,8$.
In the first case ($b=0$): (i) all methods exhibit negligible bias; (ii) $\fb$ and $\csb$ reduce MSE by 42\% and 20\%, respectively, compared to $\nb$; (iii) all methods effectively control the type I error rate; and (iv) $\fb$ and $\csb$ increase power by 46\% and 45\%, respectively, compared to $\nb$.
In the following eight cases ($b = 1, \ldots, 8$): (i) $\fb$ exhibits a large bias, approximately 125\%-203\% of its standard deviation (SD). 
The absolute bias of $\fb$ decreases with $b$ when $b \geq 3$ because large $b$ values increase $\bbV\{Y(0) \mid X=x, A=0, S=0\}$, causing $\fb$ to down-weight ECs with small $\hat{r}(X_i)$ in \eqref{eq:fb}.
$\csb$ performs better at bias control, with bias ranging from 0\%-22\% of its SD; (ii) compared to $\nb$, $\fb$ increases MSE by up to 454\%, and $\csb$ decreases MSE by 13\%-16\% (except when $b = 1,2$, where MSE increases by 1\%-18\%). (iii) In line with Theorem \ref{thm:frt}, all methods control the type I error rate well; (iv) compared to $\nb$, $\fb$ decreases power by up to 51\%. In contrast, $\csb$ increases power by 13\%-36\% (except when $b = 2,3,4$, where power decreases by 7\%-20\%).
In challenging cases where $0 < b \leq 4$, the efficiency loss of $\csb$ occurs because small biases make it hard to distinguish biased ECs from unbiased ECs. 
Such loss is inevitable when aiming to gain efficiency in scenarios without hidden bias, a phenomenon known in the transfer learning literature as the cost of transferability detection \citep{cai2024semi}. This phenomenon also occurs for other data integration estimators under hidden bias (see Figure 2 in \citet{yang2023elastic}, Figure 4 in \citet{oberst2022understanding}, and Figure 2 in \citet{lin2024data}).
% selection
Finally, we examine the selection performance of $\csb$. 
We do not expect $\csb$ to perfectly separate biased ECs from unbiased ones due to (i) the small sample size of randomized controls and (ii) finite sample noise.
As shown in Figure \ref{fig:sim_const_ne50_main}, $\csb$ discards biased ECs and some unbiased ones that aren't sufficiently similar to randomized controls, demonstrating satisfactory selection performance.

\section{Real data application}
\label{sec:rd}

%\subsection{The CALGB 9633 and NCDB data}

\begin{table*}[ht]
\centering
\caption{\label{tab:rd}Analysis results for CALGB 9633 + NCDB.}
\centering
\begin{threeparttable}
\begin{tabular}[t]{crrccrr}
\toprule
\textbf{Method} & \textbf{Est} & \textbf{SE} & \textbf{CI} & \textbf{Asym $p$} & \textbf{Exact $p$} & \textbf{\#EC}\\
\midrule
No Borrow (Dif-in-Means) & 0.135 & 0.072 & (-0.007, 0.276) & 0.062 & 0.060 & 0\\
No Borrow (AIPW) & 0.142 & 0.074 & (-0.003, 0.286) & 0.055 & 0.051 & 0\\
Full Borrow & 0.241 & 0.061 & (0.122, 0.361) & $<$0.001 & 0.031 & 335\\
Conformal Selective Borrow & 0.138 & 0.058 & (0.024, 0.252) & 0.018 & 0.046 & 264\\
\bottomrule
\end{tabular}
\begin{tablenotes}
\item {\footnotesize ``Est'' is the estimate. ``SE'', ``CI'', and ``Asym $p$'' are the asymptotic standard error, confidence interval, and $p$-value, respectively. ``Exact $p$'' is the exact $p$-value. ``\#EC'' is the number of borrowed ECs.}
\end{tablenotes}
\end{threeparttable}
\end{table*}

% intro
%\textbf{Description.}
\textbf{The CALGB 9633 and NCDB data.} We apply the proposed methods to an RCT conducted by the Cancer and Leukemia Group B (CALGB), known as CALGB 9633, which investigated the treatment effect of adjuvant chemotherapy in patients with stage IB non-small-cell lung cancer \citep{strauss2008adjuvant}.
In CALGB 9633 ($S=1$), $\nt=167$ patients were randomized to adjuvant chemotherapy ($A=1$), and $\nc=168$ were randomized to observation ($A=0$). 
We extract data for 11,700 patients from the National Cancer Database (NCDB) as the EC sample ($A=0,S=0$) to improve CALGB 9633's statistical efficiency.
The NCDB is a clinical oncology database sourced from hospital registry data, jointly run by the American Cancer Society and the American College of Surgeons, covering 70\% of U.S. cancer cases.

% RMST
\textbf{RMST and pseudo-observations} We use the \textit{Restricted Mean Survival Time} (RMST), $Y = \min(T, t^*)$, as the primary endpoint, where $T$ represents the survival time and $t^*$ is the truncation time. RMST measures survival time up to a clinically relevant truncation point and serves as a compelling alternative to the hazard ratio when the proportional hazards assumption is violated \citep{hernan2010hazards}.
We consider the difference in 3-year RMST between the treatment and control groups for the RCT population
$
\tau = E\{Y(1) - Y(0) \mid S = 1\}
$
as the estimand, where $Y(a)=\min\{T(a),3\}$ and $T(a)$ is the potential survival time, $a=0,1$.
Five baseline covariates in CALGB 9633 and NCDB are considered: sex, age, race, histology, and tumor size.
% pseudo Y
The censoring rates of $T$ in CALGB 9633 and NCDB are 42\% and 48\%, respectively. We use a ``once-for-all" approach to transform right-censored survival times into \textit{pseudo-observations} for RMST, allowing standard causal inference methods as if outcomes were non-censored \citep{andersen2003generalised,overgaard2017asymptotic}. To address covariate-dependent censoring, we stratified by sex, race, and histology, applying transformations separately within each dataset \citep{andersen2010pseudo}. The stratified Kaplan–Meier estimator is used to estimate survival functions, with pseudo-observations generated via the jackknife method, as implemented in the \texttt{R} package \texttt{eventglm} \citep{sachs2022event}. %More details are in the Supplement Material. 
We treat the pseudo-observations for 3-year RMST as the outcome hereafter.
More details about the real data are provided in Appendix Section \ref{sec:addrd}.

%\subsection{Data analysis}

% method
\textbf{Data analysis.} We apply $\nb$, $\fb$, and $\csb$ to estimate the ATE and perform FRTs. 
For comparison, we also apply $\nb$ without covariate adjustment, i.e., difference-in-means estimator.
In addition to the proposed exact $p$-value, we also compute the standard error, confidence interval, and $p$-value based on asymptotic inference for all approaches \citep{li2023improving}.
Since the outcome shows a high proportion of truncation at 3 years, resulting in a highly skewed distribution, we apply the conformal quantile regression \citep{romano2019conformalized} to compute the conformal score.
We use the Jackknife+ $p$-value \citep{barber2021predictive} to achieve a better balance between statistical and computational efficiency.
% result
Table \ref{tab:rd} presents the analysis results. For $\nb$ using Dif-in-Means and AIPW, asymptotic and exact $p$-values range from 0.051 to 0.062. In contrast, $\fb$ (using all 335 ECs) gives an asymptotic $p$-value of $<0.001$ and an exact $p$-value of 0.031, indicating a significantly positive ATE. Similarly, $\csb$ (using 178 ECs) shows an asymptotic $p$-value of 0.018 and an exact $p$-value of 0.046, also indicating a significantly positive ATE. The ATE estimate from $\csb$ falls between $\nb$ and $\fb$, indicating a trade-off between these two approaches.

\section{Discussion}

This paper proposes using FRT in hybrid controlled trials and introduces $\csb$ for selectively incorporating comparable ECs, mitigating hidden bias. FRT with $\csb$ maintains type I error control and improves power compared to RCT-only analysis. The proposed $\csb$ estimator with an adaptive selection threshold enhances efficiency over the $\nb$ approach.

One limitation of our procedure is that, when the bias is non-negligible yet difficult to detect, it may incur some power loss, though it still maintains valid Type I error control. This no-free-lunch limitation is acknowledged in existing papers \citep{oberst2022understanding,lin2024data}, which point out that without assuming mean exchangeability of ECs, no method can uniformly and significantly outperform RCT-only analysis across varying levels of hidden bias, although different approaches optimize the risk-reward trade-off from different perspectives. The most challenging scenarios are those where bias is non-negligible but complex to correct or difficult to detect. Our key distinctions from existing literature are twofold: (i) we prioritize exact Type I error control in small samples before seeking power gains; (ii) we optimize the risk-reward trade-off between no borrowing and full borrowing through conformal selective borrowing, motivated by real data in which some ECs are unbiased while others are not.

% other data integration
Heterogeneity among data sources is common in integration and transfer learning, often leading to bias or efficiency loss even after balancing measured confounders. While penalized bias estimation is a common solution, our work demonstrates that conformal inference provides greater stability and flexibility in finite samples. Extending this approach to tasks like developing individual treatment regimes \citep{chu2023targeted}, exploring treatment effect heterogeneity \citep{wu2022integrative}, and improving experimental design \citep{ruan2024electronic} shows great potential.

% extend sharp null
Beyond the sharp null, FRTs can test the weak null asymptotically using studentized or prepivoted statistics \citep{wu2021randomization,cohen2022gaussian}. Randomization-based confidence intervals can be constructed by inverting FRTs \citep{luo2021leveraging,zhu2023pair,fiksel2024exact}, and randomization inference can test bounded nulls and construct confidence intervals for treatment effect quantiles \citep{caughey2023randomisation}. Extending these methods to hybrid controlled trials would be valuable.

% Acknowledgements should only appear in the accepted version.
% \section*{Acknowledgements}

% \textbf{Do not} include acknowledgements in the initial version of
% the paper submitted for blind review.

% If a paper is accepted, the final camera-ready version can (and
% usually should) include acknowledgements.  Such acknowledgements
% should be placed at the end of the section, in an unnumbered section
% that does not count towards the paper page limit. Typically, this will 
% include thanks to reviewers who gave useful comments, to colleagues 
% who contributed to the ideas, and to funding agencies and corporate 
% sponsors that provided financial support.

\section*{Software and Data}

A user-friendly R package, \texttt{intFRT}, is available at: \url{https://github.com/ke-zhu/intFRT}.

\section*{Acknowledgment}

We thank the anonymous reviewers and meta-reviewers of ICML 2025 for their helpful comments, which significantly improved the manuscript. This project is supported by the Food and Drug Administration (FDA) of the U.S. Department of Health and Human Services (HHS) as part of a financial assistance award U01FD007934 totaling \$1,674,013 over two years funded by FDA/HHS. It is also supported by the National Institute On Aging of the National Institutes of Health under Award Number R01AG06688, totaling \$1,565,763 over four years. The contents are those of the authors and do not necessarily represent the official views of, nor an endorsement by, FDA/HHS, the National Institutes of Health, or the U.S. Government.

\section*{Impact Statement}

This paper presents work aimed at advancing data integration, conformal inference, and their applications in biomedical science. The potential societal impact of this research is substantial, including fostering the reliable and efficient use of real-world data, accelerating drug development processes, improving the understanding of rare diseases, and ultimately enhancing patient outcomes.

% Authors are \textbf{required} to include a statement of the potential 
% broader impact of their work, including its ethical aspects and future 
% societal consequences. This statement should be in an unnumbered 
% section at the end of the paper (co-located with Acknowledgements -- 
% the two may appear in either order, but both must be before References), 
% and does not count toward the paper page limit. In many cases, where 
% the ethical impacts and expected societal implications are those that 
% are well established when advancing the field of Machine Learning, 
% substantial discussion is not required, and a simple statement such 
% as the following will suffice:

% ``This paper presents work whose goal is to advance the field of 
% Machine Learning. There are many potential societal consequences 
% of our work, none which we feel must be specifically highlighted here.''

% The above statement can be used verbatim in such cases, but we 
% encourage authors to think about whether there is content which does 
% warrant further discussion, as this statement will be apparent if the 
% paper is later flagged for ethics review.

% In the unusual situation where you want a paper to appear in the
% references without citing it in the main text, use \nocite
%\nocite{langley00}

\bibliography{ref}
\bibliographystyle{icml2025}

%%%%%%%%%%%%%%%%%%%%%%%%%%%%%%%%%%%%%%%%%%%%%%%%%%%%%%%%%%%%%%%%%%%%%%%%%%%%%%%
%%%%%%%%%%%%%%%%%%%%%%%%%%%%%%%%%%%%%%%%%%%%%%%%%%%%%%%%%%%%%%%%%%%%%%%%%%%%%%%
% APPENDIX
%%%%%%%%%%%%%%%%%%%%%%%%%%%%%%%%%%%%%%%%%%%%%%%%%%%%%%%%%%%%%%%%%%%%%%%%%%%%%%%
%%%%%%%%%%%%%%%%%%%%%%%%%%%%%%%%%%%%%%%%%%%%%%%%%%%%%%%%%%%%%%%%%%%%%%%%%%%%%%%
\newpage
\appendix
\onecolumn

% \section{Related work for randomization inference and conformal inference}
% \label{sec:rel}

\section{Additional conformal \textit{p}-values}

\textbf{Full conformal \textit{p}-value.} Full conformal inference \citep{vovk2005algorithmic} fully utilizes all data in $\cC$ for both training and calibration.
We can still use the absolute residual as the score function: $s_i=|Y_i-\hf_j(X_i)|$ for $ i\in\cC$ and $s_j=|Y_j-\hf_j(X_j)|$,
% \begin{gather*}
% s_i=|Y_i-\hf_j(X_i)|,\quad i\in\cC,\\
% s_j=|Y_j-\hf_j(X_j)|,
% \end{gather*}
where $\hf_j(x)$ is a prediction model fitted by the augmented set $\cC\cup\{j\}$.
To measure the extremeness of observing $s_j$ under the exchangeability, we define the full conformal $p$-value as the proportion of the elements in $\{s_i\}_{i\in\cC}$ that are larger than or equal to $s_j$, that is, $p_{j}^\full= \{\sum_{i\in\cC}\bbI(s_i\geq s_j)+1\}/(|\cC|+1)$.
% $$
% p_{j}^\full= \frac{\sum_{i\in\cC}\bbI(s_i\geq s_j)+1}{|\cC|+1},
% $$

\begin{proposition}
\label{prop:fcf}
For $j\in\cE$, suppose that $(X_j,Y_j)$ and $\{(X_i,Y_i)\}_{i\in\cC}$ are exchangeable.
For $\gamma\in(0,1)$, we have 
$
\bbP(p_{j}^\full\leq \gamma)\leq\gamma.
$
If $s_j$ and $\{s_i\}_{i\in\cC}$ have distinct values, we have
$
\bbP(p_{j}^\full\leq \gamma)=\big\{\left\lfloor\gamma\left(|\cC|+1\right)\right\rfloor\big\}\,\big/\,(|\cC|+1)>\gamma - 1/(|\cC|+1).
$
% For an EC $j\in\cE$, suppose that $(X_j,Y_j)$ and the randomized controls $\{(X_i,Y_i)\}_{i\in\cC}$ are exchangeable. For $\gamma\in(0,1)$, we have 
% $
% \bbP(p_{j}^\full\leq \gamma)\leq\gamma.
% $
% Further assuming $s_j$ and $\{s_i\}_{i\in\cC}$ have distinct values, we have
% $
% \bbP(p_{j}^\full\leq \gamma)=\big\{\left\lfloor\gamma\left(|\cC|+1\right)\right\rfloor\big\}\,\big/\,(|\cC|+1)>\gamma - 1/(|\cC|+1).
% $
% $$
% \bbP(p_{j}^\full\leq \gamma)=\frac{\left\lfloor\gamma\left(|\cC|+1\right)\right\rfloor}{|\cC|+1}>\gamma - \frac{1}{|\cC|+1}.
% $$
\end{proposition}
To compute full conformal $p$-values for all ECs $j\in\cE$, the prediction model must be refit $\ne$ times, which is time-consuming for large EC samples.

\textbf{Jackknife+ \textit{p}-value.} Jackknife+ $p$-values \citep{barber2021predictive} is a special case of CV+ with $K=|\cC|$.
We use the leave-one-out
training set $\cC \setminus\{i\}$ to fit prediction models $\hf_{-i}(x)$ and use the absolute residual as the score function:
$s_i=|Y_i-\hf_{-i}(X_i)|$ and $\sji=|Y_j-\hf_{-i}(X_j)|$ for $i\in\cC$.
% \begin{gather*}
% s_i=|Y_i-\hf_{-i}(X_i)|,\quad i\in\cC,\\
% \sji=|Y_j-\hf_{-i}(X_j)|, \quad i\in\cC.
% \end{gather*}
We define the Jackknife+ $p$-value as the proportion of $\{s_i\}_{i\in\cC}$ that are larger than the corresponding $\{\sji\}_{i\in\cC}$, that is, $p_{j}^\jp= \{\sum_{i\in\cC}\bbI(s_i\geq \sji)+1\}/(|\cC|+1)$.
% $$
% p_{j}^\jp= \frac{\sum_{i\in\cC}\bbI(s_i\geq \sji)+1}{|\cC|+1}.
% $$

\begin{proposition}
\label{prop:jp}
For $j\in\cE$, suppose that $(X_j,Y_j)$ and $\{(X_i,Y_i)\}_{i\in\cC}$ are exchangeable. For $\gamma\in(0,1)$, we have 
$
\bbP(p_{j}^\jp\leq \gamma)\leq2\gamma-1/(|\cC|+1)<2\gamma.
$
% $$
% \bbP(p_{j}^\jp\leq \gamma)\leq2\gamma-\frac{1}{|\cC|+1}<2\gamma.
% $$
\end{proposition}

\begin{remark}
    The factor of 2 cannot be reduced without further assumptions, as shown by pathological cases in \citet{barber2021predictive}, though the empirical error rate is close to $\gamma$.
\end{remark}

\section{Proofs}
\label{sec:proof}

\subsection{Proof of Theorem \ref{thm:frt}}

\begin{proof}[Proof of Theorem \ref{thm:frt}]
Under $H_0$, the imputed potential outcomes are the same as the true potential outcomes. Thus, the distribution of $T^*\equiv T(\A^*)$ is the same as that of $T\equiv T(\A)$. 
With simplified notations, we have 
$$
\bbPA(p^{\rm FRT}\leq\alpha)
\;=\;
\bbPA\left\{\bbPAs\left(
T^*\geq T
\right)\leq\alpha\right\}.
$$
In a finite sample, $\A$ can take only a finite set of values, which implies that $T$ must also take on a finite set of values.
Suppose these values are 
$$
T_1>\ldots>T_m>\ldots>T_M,
$$
and
$$
\bbPA(T=T_m)
\;=\;
\bbPAs(T^*=T_m)\;=\;
\alpha_m,\quad m=1,\ldots,M.
$$
For $T\in\{T_1,\ldots,T_{M}\}$, we have $\alpha_1\leq\bbPAs\left(
T^*\geq T
\right)\leq \sum_{m=1}^M\alpha_m=1$. 
If $0<\alpha<\alpha_1$, we have
$$
\bbPA(p^{\rm FRT}\leq\alpha)
\;=\;
\bbPA\left\{\bbPAs\left(
T^*\geq T
\right)\leq\alpha\right\} 
\;=\;
0
\;\leq\;
\alpha.
$$
If $\alpha_1\leq\alpha< 1$, $\exists\tilde{M}\in\{1,\ldots,M-1\}$, such that $\sum_{m=1}^{\tilde{M}}\alpha_m\leq\alpha$ and $\sum_{m=1}^{\tilde{M}+1}\alpha_m>\alpha$.
Then, we have
\begin{align*}
\bbPA(p^{\rm FRT}\leq\alpha)
\;=\;
\bbPA\left\{\bbPAs\left(T^*\geq T\right)\leq\alpha\right\} 
\;=\;
\bbPA\left\{T\in\{T_1,\ldots,T_{\tilde{M}}\}\right\}
\;=\;
\sum_{m=1}^{\tilde{M}}\alpha_m
\;\leq\;
\alpha.
\end{align*}

If $T(\A)$ takes distinct values for different $\A\in\cA$, $p^{\rm FRT}$ is uniformly distributed:
$$
\bbPA\left(p^{\rm FRT}=\frac{a}{|\cA|}\right)=\frac{1}{|\cA|}, \quad a=1, \ldots, |\cA|. 
$$
Thus, we have
$$
\bbPA(p^{\rm FRT}\leq\alpha)
=\frac{\lfloor\alpha|\cA|\rfloor}{|\cA|}
>\frac{\alpha|\cA|-1}{|\cA|}
=\alpha - \frac{1}{|\cA|}.
$$
\end{proof}

\begin{remark}
If $T$ is a continuous random variable, suppose its distribution function is $F(t) = P(T\leq t)$, then the proof could be simplified as
\begin{align*}
\bbPA\left\{\bbPAs\left(
T^*\geq T
\right)\leq\alpha\right\}
&=P\left\{1-F(T)\leq\alpha\right\}\\
&=P\left\{T\geq F^{-1}(1-\alpha)\right\}\\
&=1-F\{F^{-1}(1-\alpha)\}\\
&=\alpha.
\end{align*}
However, $T$ is discrete with finite values, and we provide a rigorous proof in the finite-sample setting.
\end{remark}

\subsection{Proof of Theorem \ref{thm:power}}

We invoke two lemmas from the Supplementary Material of \citet{puelz2022graph}.

\begin{lemma}[Lemma 5 in \citet{puelz2022graph}]
\label{lm:T2}
Suppose Assumptions (b) and (c) in Theorem \ref{thm:power} hold, for some $r\in (0.5,1+O(\log^{-1}M))$, we have  
$$
\bbE(\pF{1}(\pq) - \pF{1}(\hq))\geq-O(M^{-r}),
$$
\end{lemma}

\begin{lemma}[Lemma 4 in \citet{puelz2022graph}]
\label{lm:T3}
Suppose Assumption (a) of Theorem \ref{thm:power} holds, for any $0<\iota<0.5$ and large enough $M$, we have
$$
\bbE(\hF{1}(z) - \pF{1}(z))=O(M^{-0.5+\iota}),  \quad \text{for any } z\in\bbR.
$$
\end{lemma}

\begin{proof}[Proof of Theorem \ref{thm:power}]
Let $\hq = \hF{0}^{-1}(1-\alpha)$ and  $\pq = \pF{0}^{-1}(1-\alpha)$. Thus, we have 
\begin{align}
\psi_{N,M}&=1-\hF{1}\big(\hF{0}^{-1}(1-\alpha)\big) \nonumber\\
&=1-\hF{1}\big(\hq\big) \nonumber\\
&=
\underbrace{1-\pF{1}(\pq)}_{T_1} + 
\underbrace{\pF{1}(\pq) - \pF{1}(\hq)}_{T_2} + 
\underbrace{\pF{1}(\hq) - \hF{1}(\hq)}_{T_3}.
\end{align}
By Assumptions (b) and (c), we have
$$
T_1 = 1-\pF{1}\big(\pF{0}^{-1}(1-\alpha)\big)=1-F\left(F^{-1}(1-\alpha)-\tau / \sigma_N\right).
$$
Combined with Lemmas \ref{lm:T2} and \ref{lm:T3}, we have
$$
\bbE(\psi_{N,M})\geq
1-F\left(F^{-1}(1-\alpha)-\tau / \sigma_N\right)
-O(M^{-r})
-O(M^{-0.5+\iota}).
$$
The result follows from that $r>0.5>0.5-\iota>0$.
\end{proof}

\subsection{Proof of Proposition \ref{prop:fcf}}

\begin{proof}[Proof of Proposition \ref{prop:fcf}]
Since the calibration set $(X_i,Y_j)_{i\in\cC}$ and external control $(X_j,Y_j)$ are exchangeable, we have $(s_i)_{i\in\cC}$ and $s_j$ are exchangeable. 
Thus, we have
\begin{align*}
\bbP(p_{j}^\full\leq \gamma)=&
\bbP\left(\frac{\sum_{i\in\cC}\bbI(s_i\geq s_j)+1}{|\cC|+1}\leq \gamma\right)\\
%=&\bbP\left(\sum_{i\in\cC}\bbI(s_i\geq s_j)+1\leq \left\lfloor\gamma\left(|\cC|+1\right)\right\rfloor\right)\\
\leq&\frac{\left\lfloor\gamma\left(|\cC|+1\right)\right\rfloor}{|\cC|+1}\\
\leq& \gamma,
\end{align*}
where the first inequality is due to exchangeability and the possibility of ties in $(s_i)_{i\in\cC}$ and $s_j$.

If $s_j$ and $\{s_i\}_{i\in\cC}$ have distinct values, $p_{j}^\full$ is uniformly distributed due to exchangeability. That is,
$$
\bbP\left(p_{j}^\full=\frac{a}{|\cC|+1}\right)=\frac{1}{|\cC|+1}, \quad a=1, \ldots, |\cC|+1. 
$$
Thus, we have
$$
\bbP(p_{j}^\full\leq \gamma)=
\frac{\left\lfloor\gamma\left(|\cC|+1\right)\right\rfloor}{|\cC|+1}
>\frac{\gamma\left(|\cC|+1\right)-1}{|\cC|+1}=\gamma - \frac{1}{|\cC|+1}.
$$
\end{proof}

\subsection{Proof of Proposition \ref{prop:scf}}
\begin{proof}[Proof of Proposition \ref{prop:scf}]
Since the calibration set $(X_i,Y_j)_{i\in\cC}$ and external control $(X_j,Y_j)$ are exchangeable, we have $(s_i)_{i\in\cC_1}$ and $s_j$ are exchangeable. 
Thus, we have
\begin{align*}
\bbP(p_{j}^\sp \leq \gamma)=&
\bbP\left(\frac{\sum_{i\in\cC_1}\bbI(s_i\geq s_j)+1}{|\cC_1|+1}\leq \gamma\right)\\
\leq&\frac{\lfloor\gamma(|\cC_1|+1)\rfloor}{|\cC_1|+1}\\
\leq& \gamma,
\end{align*}
where the first inequality is due to exchangeability and the possibility of ties in $(s_i)_{i\in\cC_1}$ and $s_j$.

If $s_j$ and $\{s_i\}_{i\in\cC_1}$ have distinct values, $p_{j}^\sp$ is uniformly distributed due to exchangeability. That is,
$$
\bbP\left(p_{j}^\sp=\frac{a}{|\cC_1|+1}\right)=\frac{1}{|\cC_1|+1}, \quad a=1, \ldots, |\cC_1|+1. 
$$
Thus, we have
$$
\bbP(p_{j}^\sp\leq \gamma)=
\frac{\lfloor\gamma(|\cC_1|+1)\rfloor}{|\cC_1|+1}
>\frac{\gamma(|\cC_1|+1)-1}{|\cC_1|+1}=\gamma - \frac{1}{|\cC_1|+1}.
$$
\end{proof}

\subsection{Proof of Proposition \ref{prop:jp}}

\begin{lemma}
\label{lm:jack_bound}
Consider a matrix $R\in\bbR^{(n+1)\times(n+1)}$ with elements $R_{ij}$. Define the set
$$
\cS=\left\{j\in\{1,\ldots,n+1\}:\sum_{i=1}^{n+1}\bbI(R_{ij}< R_{ji})\geq(1-\gamma)(n+1)\right\}, \quad \gamma\in(0,1).
$$
Then, we have
$$
s\leq2\gamma(n+1)-1 <2\gamma(n+1),
$$
where $s=|\cS|$.
\end{lemma}
\begin{proof}
Since
$$
\sum_{i=1}^{n+1}\bbI(R_{ij}< R_{ji})\geq(1-\gamma)(n+1)
\quad\Leftrightarrow\quad
\sum_{i=1}^{n+1}\bbI(R_{ij}\geq R_{ji})\leq\gamma(n+1),
$$
by summing over all $j\in\cS$, we have
\begin{align*}
\sum_{j\in\cS}\sum_{i=1}^{n+1}\bbI(R_{ij}\geq R_{ji})\leq s\gamma(n+1).
\end{align*}
For $i\neq j$, since $\bbI(R_{ij}\geq R_{ji})+\bbI(R_{ji}\geq R_{ij})\geq 1$, we have
\begin{align*}
\sum_{j\in\cS}\sum_{i\in\cS}\bbI(R_{ij}\geq R_{ji})=&\sum_{j\in\cS}\sum_{i\in\cS, i\neq j}\bbI(R_{ij}\geq R_{ji})+s \\
\geq& \frac{s(s-1)}{2}+s.
\end{align*}
By combining these two inequalities, we obtain
\begin{align*}
\frac{s(s-1)}{2}+s
\leq&
\sum_{j\in\cS}\sum_{i\in\cS}\bbI(R_{ij}\geq R_{ji})\\
\leq&
\sum_{j\in\cS}\sum_{i=1}^{n+1}\bbI(R_{ij}\geq R_{ji})\\
\leq& s\gamma(n+1).
\end{align*}
Thus, we have
$$
\frac{(s-1)}{2}+1\leq\gamma(n+1) \quad\Rightarrow \quad 
s\leq2\gamma(n+1)-1 <2\gamma(n+1).
$$
\end{proof}

\begin{proof}[Proof of Proposition \ref{prop:jp}]
For $\i,\j\in \cC \cup\{j\}$, we define
$$
R_{\i\j}= \begin{cases}+\infty & \i=\j, \\ \left|Y_\i-\hf_{-(\i, \j)}\left(X_\i\right)\right| & \i \neq \j,\end{cases}
$$
where $\hf_{-(\i, \j)}$ is a prediction model fitted by the leave-two-out augmented set $(\cC \cup\{j\})\setminus \{\i,\j\}$.
For $i\in\cC$, since $(\cC \cup\{j\})\setminus \{i,j\}=\cC\setminus \{i\}$, we have $\hf_{-i}(x)=\hf_{-(i,j)}(x)$, thereby,
\begin{gather*}
s_i=|Y_i-\hf_{-i}(X_i)|=R_{ij},\\
\sji=|Y_j-\hf_{-i}(X_j)|=R_{ji}.
\end{gather*}
Thus, we have
\begin{align*}
\bbP(p_{j}^\jp\leq \gamma) = &
\bbP\left( \frac{\sum_{i\in\cC}\bbI(s_i\geq \sji)+1}{|\cC|+1}\leq \gamma\right) \\
=&\bbP\left( \frac{\sum_{i\in\cC\cup\{j\}}\bbI(R_{ij}\geq R_{ji})}{|\cC|+1}\leq \gamma\right) \\
=&\bbP\left( \sum_{i\in\cC\cup\{j\}}\bbI(R_{ij}< R_{ji})\geq (1-\gamma)(|\cC|+1)\right) \\
\leq& 2\gamma-\frac{1}{|\cC|+1}\\
<&2\gamma,
\end{align*}
where first inequality is due to exchangeability and Lemma \ref{lm:jack_bound}.
\end{proof}

\subsection{Proof of Proposition \ref{prop:cvp}}

\begin{lemma}
\label{lm:cv_bound}
Suppose $m = n/K$ is an integer, and the $n + m$ units are evenly divided into $K + 1$ sets, denoted by $\cC_1,\ldots,\cC_{K+1}$.
Consider a matrix $R \in \bbR^{(n+m) \times (n+m)}$ with elements $R_{ij} = R_{ji}$ if $i$ and $j$ belong to the same set. Define the set
$$
\cS = \left\{ j \in \{1, \ldots, n+m\} : \sum_{i=1}^{n+m} \bbI(R_{ij} < R_{ji}) \geq (1-\gamma)(n+1) \right\}, \quad \gamma \in (0, 1).
$$
Then, we have
$$
s \leq 2\gamma(n+1) + m-2,
$$
where $s=|\cS|$.
\end{lemma}
\begin{proof}
For $j\in\cS$, by definition, we have
$$
\sum_{i=1}^{n+m}\bbI(R_{ij}\geq R_{ji})\leq(n+m)-(1-\gamma)(n+1).
$$
Since $R_{ij} = R_{ji}$ if $i$ and $j$ belong to the same set, we have
\begin{align*}
\sum_{i=1}^{n+m}\bbI(R_{ij}\geq R_{ji})=&
\sum_{i\notin\cC_{k(j)}}\bbI(R_{ij}\geq R_{ji})
+\sum_{i\in\cC_{k(j)}}\bbI(R_{ij}\geq R_{ji})
\\
=&\sum_{i\notin\cC_{k(j)}}\bbI(R_{ij}\geq R_{ji})+m,
\end{align*}
where $\cC_{k(j)}$ is the set containing unit $j$. 
Thus, we have
\begin{align*}
\sum_{i\notin\cC_{k(j)}}\bbI(R_{ij}\geq R_{ji})\leq&
(n+m)-(1-\gamma)(n+1)-m\\
=&\gamma(n+1)-1.   
\end{align*}
By summing over all $j\in\cS$, we have
\begin{equation}
\label{eq:cv1}
\sum_{j\in\cS}\sum_{i\notin\cC_{k(j)}}\bbI(R_{ij}\geq R_{ji})\leq s\{\gamma(n+1)-1\}.
\end{equation}

On the other hand, for $i\neq j$, since $\bbI(R_{ij}\geq R_{ji})+\bbI(R_{ji}\geq R_{ij})\geq 1$, we have
$$
\sum_{j\in\cS}\sum_{i\in\cS, i\neq j}\bbI(R_{ij}\geq R_{ji})\geq \frac{s(s-1)}{2}.
$$
Since $R_{ij} = R_{ji}$ if $i$ and $j$ belong to the same set, we have
\begin{align*}
\sum_{j\in\cS}\sum_{i\in\cS, i\neq j}\bbI(R_{ij}\geq R_{ji}) =&\sum_{j\in\cS}\sum_{i\in\cS, i\notin \cC_{k(j)}}\bbI(R_{ij}\geq R_{ji}) 
+\sum_{j\in\cS}\sum_{i\in\cS, i\in \cC_{k(j)},i\neq j}\bbI(R_{ij}\geq R_{ji})\\
=&\sum_{j\in\cS}\sum_{i\in\cS, i\notin \cC_{k(j)}}\bbI(R_{ij}\geq R_{ji}) +\sum_{k=1}^{K+1} \frac{s_{k}(s_{k}-1)}{2},
\end{align*}
where $s_k=|\cC_k\cap\cS|$. Thus, we have
\begin{equation}
\label{eq:cv2}
\sum_{j\in\cS}\sum_{i\in\cS, i\notin \cC_{k(j)}}\bbI(R_{ij}\geq R_{ji}) \geq \frac{s(s-1)}{2}-\sum_{k=1}^{K+1} \frac{s_{k}(s_{k}-1)}{2}.
\end{equation}
By combining \eqref{eq:cv1} and \eqref{eq:cv2}, we have
\begin{align*}
\frac{s(s-1)}{2}-\sum_{k=1}^{K+1} \frac{s_{k}(s_{k}-1)}{2}
\leq&
\sum_{j\in\cS}\sum_{i\in\cS, i\notin \cC_{k(j)}}\bbI(R_{ij}\geq R_{ji})\\
\leq&
\sum_{j\in\cS}\sum_{i\notin\cC_{k(j)}}\bbI(R_{ij}\geq R_{ji})\\
\leq& s\{\gamma(n+1)-1\}.
\end{align*}
Since $s_k\leq m$, we have
$$
\sum_{k=1}^{K+1}\frac{s_{k}(s_{k}-1)}{2}\leq \frac{s(m-1)}{2}.
$$
Thus, we have
\begin{align*}
s\leq 2\gamma(n+1) + m-2.
\end{align*}

\end{proof}

\begin{proof}[Proof of Proposition \ref{prop:cvp}]
We consider $m=|\cC|/K$ is an integer for simplicity. Let $\cC_{K+1}$ contain $j$ and other $m-1$ hypothetical points.
For $\i,\j\in \cup_{k=1}^{K+1}\cC_{k}$, we define
$$
R_{\i\j}= \begin{cases}+\infty & k(\i)=k(\j), \\ \left|Y_\i-\hf_{-(\cC_{k(\i)}, \cC_{k(\j)})}\left(X_\i\right)\right| & k(\i) \neq k(\j),\end{cases}
$$
where $\hf_{-(\cC_{k(\i)}, \cC_{k(\j)})}$ is a prediction model fitted by the leave-two-set-out augmented set $(\cup_{k=1}^{K+1}\cC_{k})\setminus (\cC_{k(\i)}\cup\cC_{k(\j)})$.
Since $\cC=\cup_{k=1}^{K}\cC_{k}$ and $\cC_{k(j)}=\cC_{K+1}$, we have $(\cup_{k=1}^{K+1}\cC_{k})\setminus (\cC_{k(i)}\cup\cC_{k(j)})=\cC\setminus \cC_{k(i)}$ for $i\in\cC$.
Thus, for $i\in\cC$, we have $\hf_{-\cC_{k(i)}}(x)=\hf_{-(\cC_{k(i)}, \cC_{k(j)})}(x)$, thereby,
\begin{gather*}
s_i=|Y_i-\hf_{-\cC_{k(i)}}(X_i)|=R_{ij},\\
\sji=|Y_j-\hf_{-\cC_{k(i)}}(X_j)|=R_{ji}.
\end{gather*}
Thus, we have
\begin{align*}
\bbP(p_{j}^\cvp\leq \gamma) = &
\bbP\left( \frac{\sum_{i\in\cC}\bbI(s_i\geq \sji)+1}{|\cC|+1}\leq \gamma\right) \\
=&\bbP\left( \frac{\sum_{i\in\cC\cup\{j\}}\bbI(R_{ij}\geq R_{ji})}{|\cC|+1}\leq \gamma\right) \\
=&\bbP\left( \sum_{i\in\cC\cup\{j\}}\bbI(R_{ij}< R_{ji})\geq (1-\gamma)(|\cC|+1)\right) \\
\leq&
\bbP\left( \sum_{i\in\cup_{k=1}^{K+1}\cC_{k}}\bbI(R_{ij}< R_{ji})\geq (1-\gamma)(|\cC|+1)\right)\\
\leq&\frac{2\gamma(|\cC|+1) + m-2}{|\cC|+m}\\
\leq&2\gamma+\frac{(1-2\gamma)(m-1) -1}{|\cC|+m}\\
\leq&2\gamma + \frac{1-K/|\cC|}{K+1},
\end{align*}
where the second inequality is due to exchangeability and Lemma \ref{lm:cv_bound}.
\end{proof}

\subsection{Proof of Theorem \ref{thm:mse2}}

\begin{proof}[Proof of Theorem \ref{thm:mse2}]
Since $\eg$ is a centered sub-Gaussian variable with parameter $\pg$, we have $\eg-\eo$ as a centered sub-Gaussian variable with parameter $2\pcap$, where $\pcap=\max_{\gamma\in\Gamma}\pg$. Moreover, we have $(\eg-\eo)^2-\kg^2$ is a centered sub-exponential variable with parameters $(c_1\pcap^2,c_1\pcap^2)$, where $c_1$ is a constant. 
By $\htg-\hto=(\dg-\don)+(\eg-\eo)$ and using the concentration inequalities for sub-Gaussian and sub-exponential variables \citep{wainwright2019high}, it follows that, with probability at least $1-4\iota$,
\begin{align}
\label{eq-mse1}
&\max_{\gamma\in\Gamma}|(\htg-\hto)^2-(\dg-\don)^2-\kg^2| \nonumber\\
=&\max_{\gamma\in\Gamma}| 2(\dg-\don)(\eg-\eo)+(\eg-\eo)^2-\kg^2| \nonumber\\
\leq & 8\sqrt{2}\dcap\pcap\sqrt{\log\left(|\Gamma|/\iota\right)}
+\max\left\{\sqrt{2}c_1\pcap^2\sqrt{\log\left(|\Gamma|/\iota\right)},\;2c_1\pcap^2\log\left(|\Gamma|/\iota\right)\right\},
\end{align}
where $\Delta=\max_{\gamma\in\Gamma}|\dg|$.

By \eqref{eq-mse1}, it follows that, with probability at least $1-4\iota$,
\begin{align*}
&\max_{\gamma\in\Gamma}\left|
\hmse{\gamma} -
\mse{\gamma}
\right|\\
=&\max_{\gamma\in\Gamma}\left|
(\htg - \hto)^2 - \widehat{\bbV}(\htg - \hto) + \widehat{\bbV}(\htg) 
-\dg^2-\sg^2
\right|\\
\leq&\max_{\gamma\in\Gamma}\left|
(\htg - \hto)^2 -\kg^2 + \sg^2-
\dg^2-\sg^2
\right|+
\max_{\gamma\in\Gamma} |\widehat{\bbV}(\htg - \hto)-\kg^2|
+ \max_{\gamma\in\Gamma} |\widehat{\bbV}(\htg)-\sg^2|\\
\leq &\max_{\gamma\in\Gamma}\left|
(\dg-\don)^2-\dg^2
\right|+c\dcap\pcap\sqrt{\log\left(|\Gamma|/\iota\right)}
+\max\left\{c\pcap^2\sqrt{\log\left(|\Gamma|/\iota\right)},\;c\pcap^2\log\left(|\Gamma|/\iota\right)\right\}\\
&+
\max_{\gamma\in\Gamma} |\widehat{\bbV}(\htg - \hto)-\kg^2|
+ \max_{\gamma\in\Gamma} |\widehat{\bbV}(\htg)-\sg^2|\\
\leq &c\dcap|\don|+
c\dcap\pcap\sqrt{\log\left(|\Gamma|/\iota\right)}
+\max\left\{c\pcap^2\sqrt{\log\left(|\Gamma|/\iota\right)},\;c\pcap^2\log\left(|\Gamma|/\iota\right)\right\}\\
&+\max_{\gamma\in\Gamma} |\widehat{\bbV}(\htg - \hto)-\kg^2|
+ \max_{\gamma\in\Gamma} |\widehat{\bbV}(\htg)-\sg^2|,
\end{align*}
where $c$ is a constant.
\end{proof}

\subsection{Proof of Theorem \ref{cor:mse3}}

\begin{proof}[Proof of Theorem \ref{cor:mse3}]
Since $\eg$ is a centered sub-Gaussian variable with parameter $\pg$, we have $\eg^2-\sg^2$ as a centered sub-exponential variable with parameter $(c_2\pcap^2,c_2\pcap^2)$, where $c_2$ is a constant. 
By $\htg-\tau=\dg+\eg$ and using the concentration inequalities for sub-Gaussian and sub-exponential variables \citep{wainwright2019high}, it follows that, with probability at least $1-4\iota$,
\begin{align}
\label{eq-mse2}
&\max_{\gamma\in\Gamma}|(\htg-\tau)^2-\dg^2-\sg^2|\nonumber\\
=&\max_{\gamma\in\Gamma}| 2\dg\eg+\eg^2-\sg^2|\nonumber\\
\leq & 2\sqrt{2}\dcap\pcap\sqrt{\log\left(|\Gamma|/\iota\right)}
+\max\left\{\sqrt{2}c_2\pcap^2\sqrt{\log\left(|\Gamma|/\iota\right)},\;2c_2\pcap^2\log\left(|\Gamma|/\iota\right)\right\}.
\end{align}

By \eqref{eq-mse1} and \eqref{eq-mse2}, it follows that, with probability at least $1-8\iota$,
\begin{align}
\label{eq-mse3}
&\max_{\gamma\in\Gamma}\left|
\hmse{\gamma} -
(\htg-\tau)^2
\right| \nonumber\\
=&\max_{\gamma\in\Gamma}\left|
(\htg - \hto)^2 - \widehat{\bbV}(\htg - \hto) + \widehat{\bbV}(\htg) -
(\htg-\tau)^2
\right|\nonumber\\
\leq&\max_{\gamma\in\Gamma}\left|
(\htg - \hto)^2 -\kg^2 + \sg^2-
(\htg-\tau)^2
\right|+
\max_{\gamma\in\Gamma} |\widehat{\bbV}(\htg - \hto)-\kg^2|
+ \max_{\gamma\in\Gamma} |\widehat{\bbV}(\htg)-\sg^2|\nonumber\\
\leq &\max_{\gamma\in\Gamma}\left|
(\dg-\don)^2-\dg^2
\right|+c\dcap\pcap\sqrt{\log\left(|\Gamma|/\iota\right)}
+\max\left\{c\pcap^2\sqrt{\log\left(|\Gamma|/\iota\right)},\;c\pcap^2\log\left(|\Gamma|/\iota\right)\right\}\nonumber\\
&+
\max_{\gamma\in\Gamma} |\widehat{\bbV}(\htg - \hto)-\kg^2|
+ \max_{\gamma\in\Gamma} |\widehat{\bbV}(\htg)-\sg^2|\nonumber\\
\leq &c\dcap|\don|+
c\dcap\pcap\sqrt{\log\left(|\Gamma|/\iota\right)}
+\max\left\{c\pcap^2\sqrt{\log\left(|\Gamma|/\iota\right)},\;c\pcap^2\log\left(|\Gamma|/\iota\right)\right\}\nonumber\\
&+\max_{\gamma\in\Gamma} |\widehat{\bbV}(\htg - \hto)-\kg^2|
+ \max_{\gamma\in\Gamma} |\widehat{\bbV}(\htg)-\sg^2|,
\end{align}
where $c$ is a constant.

Since
$$
\left|
\min_{\gamma\in\Gamma}\hmse{\gamma} -
\min_{\gamma\in\Gamma}(\htg-\tau)^2 
\right|
\leq \max_{\gamma\in\Gamma}\left|
\hmse{\gamma} -
(\htg-\tau)^2
\right|,
$$
and 
\begin{align*}
\left|
\min_{\gamma\in\Gamma}\hmse{\gamma} -
(\htau{\hg}-\tau)^2 
\right|=
\left|
\hmse{\hg} -
(\htau{\hg}-\tau)^2 
\right|
\leq
\max_{\gamma\in\Gamma}\left|
\hmse{\gamma} -
(\htg-\tau)^2
\right|,
\end{align*}
we have
$$
(\htau{\hg} - \tau)^2
-\min_{\gamma\in\Gamma} (\htau{\gamma} - \tau)^2 \leq 2\max_{\gamma\in\Gamma}\left|
\hmse{\gamma} -
(\htg-\tau)^2
\right|.
$$
The result follows from \eqref{eq-mse3}.
\end{proof}

\section{Additional simulation results}
\label{sec:addsim}

\subsection{Power curve}

\begin{figure}[t]
    \centering
    \includegraphics[width=0.6\linewidth]{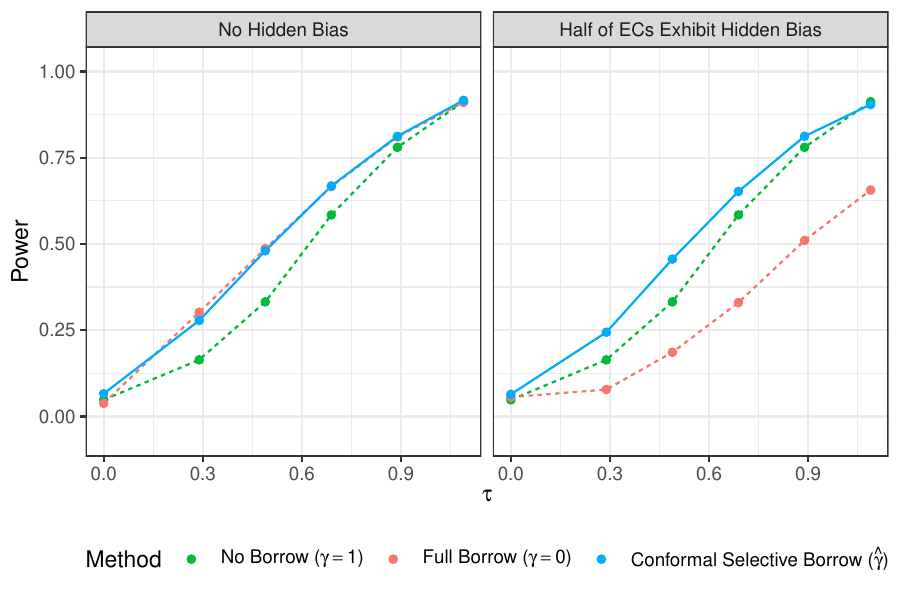}
    \caption{{Power curves when $b=0$ and $b=8$.}}
    \label{fig:sim_power}
\end{figure}

For the scenario where there is no hidden bias ($b=0$) and another where half of the ECs exhibit hidden bias with a magnitude of $b=8$, we vary $\tau$ to plot the power curve, as shown in Figure \ref{fig:sim_power}. $\csb$ outperforms $\nb$ in both cases, while $\fb$ demonstrates low power in the presence of hidden bias.

\subsection{Adaptivity of the selection threshold}

\begin{figure}[t]
    \centering
    \includegraphics[width=1\linewidth]{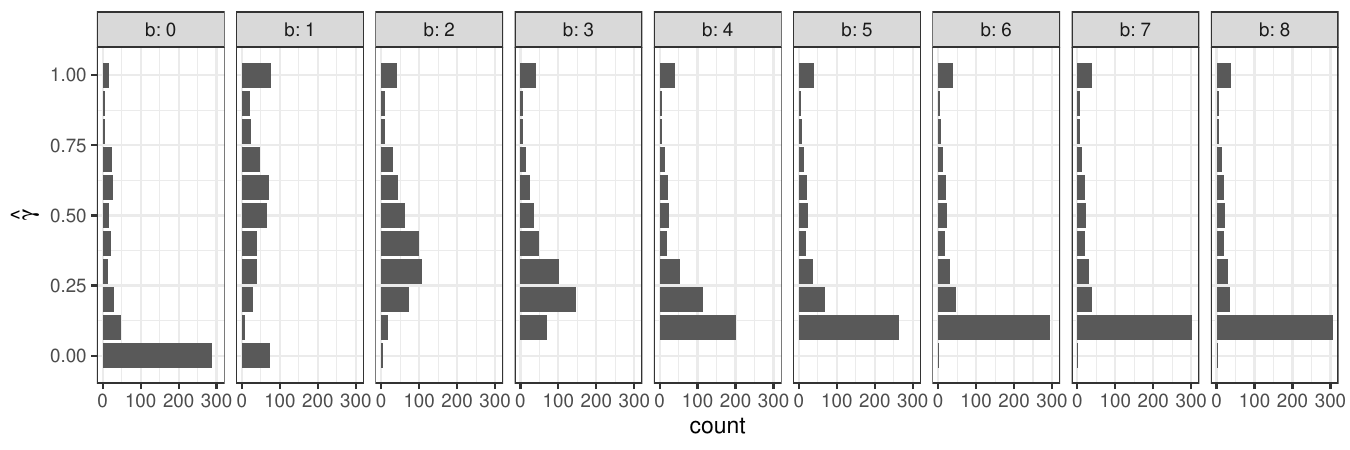}
    \caption{{$\hg$ versus $b$ when $\ne=50$.}}
    \label{fig:sim_const_ne50_adag}
\end{figure}

Figure \ref{fig:sim_const_ne50_adag} illustrates how $\hg$ changes with the magnitude of $b$:  
(i) When there is no bias ($b=0$), $\hg$ approaches 0 to borrow all ECs and maximize power; (ii) with moderate bias ($b=1, 2, 3$), where distinguishing between biased and unbiased ECs is challenging, $\hg$ increases to help discard the biased ECs; (iii) when the bias is large ($b \geq 4$), $\hg$ decreases but remains non-zero, retaining more unbiased ECs, while easily discarding the biased ones.

\subsection{Various selection thresholds}

\begin{figure}[t]
    \centering
    \includegraphics[width=1\linewidth]{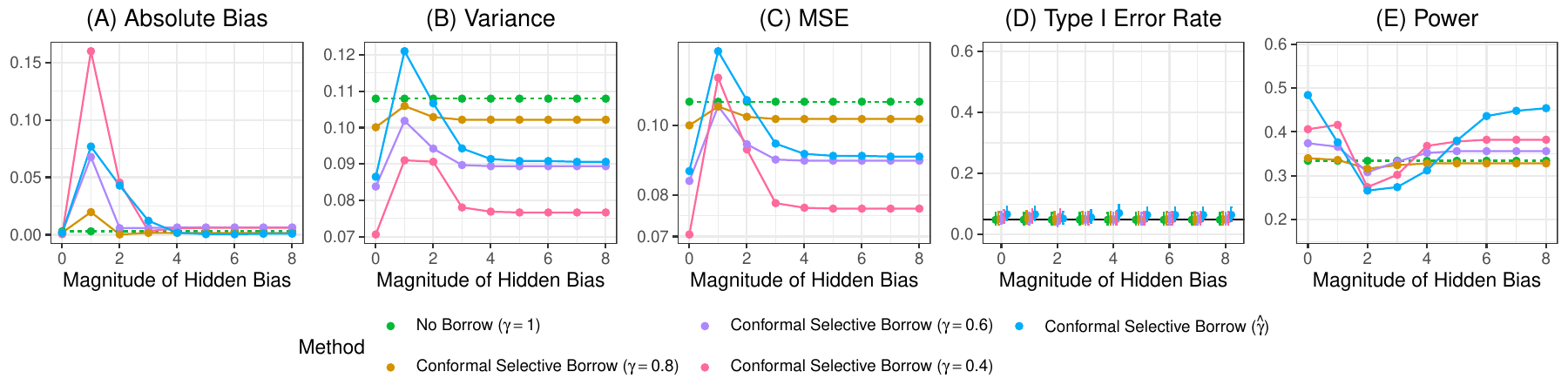}
    \caption{{Simulation results for various selection threshold $\gamma$'s when $\ne=50$.}}
    \label{fig:sim_const_ne50_difg}
\end{figure}

% gamma
Figure \ref{fig:sim_const_ne50_difg} shows the performance of the fixed selection threshold $\gamma$ and the adaptive selection threshold $\hg$ when $\ne=50$. As discussed in Section \ref{sec:ada_t}, smaller $\gamma$ selects more ECs but risks greater bias when distinguishing between biased and unbiased ECs is difficult. This creates a power trade-off across different bias levels, similar to MSE simulation results in data integration \citep{yang2023elastic,oberst2022understanding,lin2024data}. We find that (i) $\csb$ with $\gamma = 0.6$ improves power compared to NB, except in extreme cases like $b = 2,3$, where it decreases power slightly, and (ii) $\csb$ with $\hg$ further improves power but also risks power loss in difficult scenarios. The power trade-off does not compromise the Type I error rate, which remains controlled with all selection thresholds.

\subsection{Comparison to Adaptive Lasso Selective Borrowing}

\begin{figure}[t]
    \centering
    \includegraphics[width=1\linewidth]{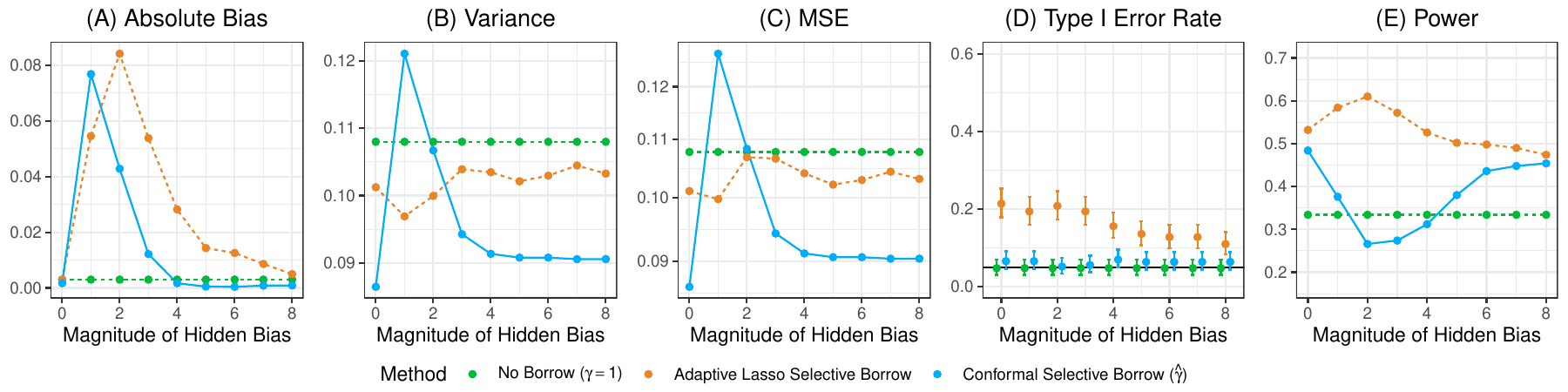}
    \caption{{Comparison of $\csb$ + FRT and $\alsb$ + asymptotic inference when $\ne=50$.}}
\label{fig:sim_const_ne50_alasso}
\end{figure}

Figure \ref{fig:sim_const_ne50_alasso} presents the simulation results for $\alsb$ with asymptotic inference. Unlike $\csb$ + FRT, $\alsb$ with asymptotic inference fails to control the type I error rate in this small sample size scenario. Additionally, $\csb$ demonstrates better estimation and selection performance in most cases.

\begin{figure}[t]
    \centering
    \includegraphics[width=1\linewidth]{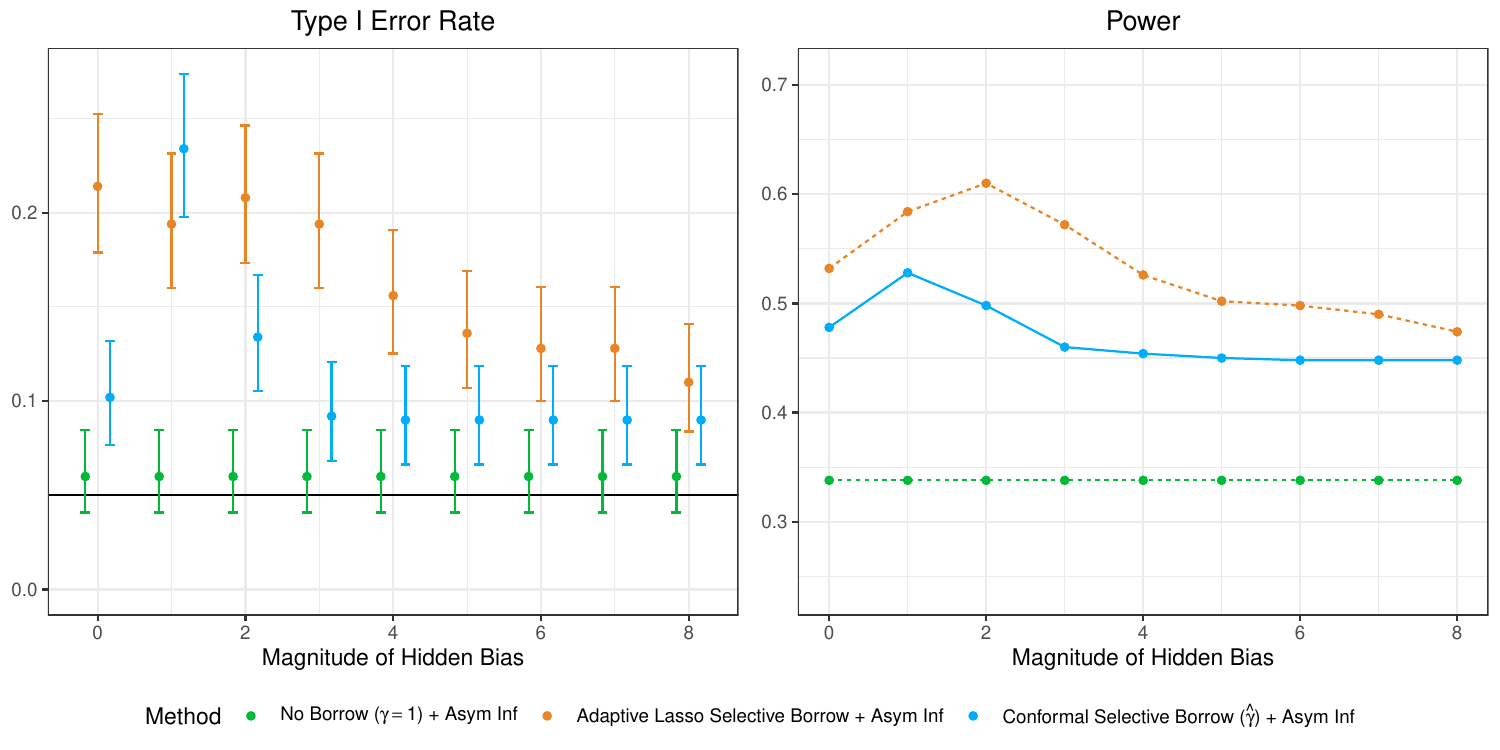}
    \caption{{Comparison of $\csb$ + asymptotic inference and $\alsb$ + asymptotic inference when $\ne=50$.}}
\label{fig:sim_csb_asym}
\end{figure}

We further compared $\csb$ with asymptotic inference to $\alsb$ with asymptotic inference. Figure \ref{fig:sim_csb_asym} shows that $\csb$+Asym Inf generally achieves better Type I error control than $\alsb$+Asym Inf, while performing comparably when $b = 1$.

We did not compare to $\alsb$ + FRT because, while $\csb$ is compatible with FRT, $\alsb$ is not readily applicable due to its computational complexity. This highlights an advantage of $\csb$ when exact finite-sample inference is desired.

\subsection{A larger sample size of ECs}

\begin{figure}[t]
    \centering
    \includegraphics[width=1\linewidth]{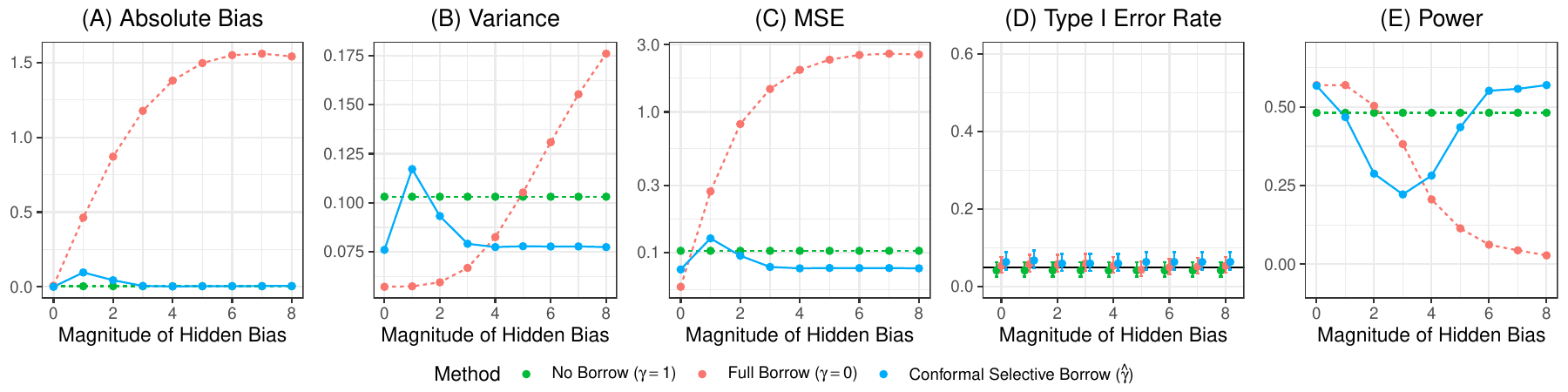}
    \caption{{Simulation results when $\ne=300$. $\alsb$'s exact $p$-value is unavailable due to computation.}}
    \label{fig:sim_const_ne300_main}
\end{figure}

\begin{figure}[t]
    \centering
    \includegraphics[width=1\linewidth]{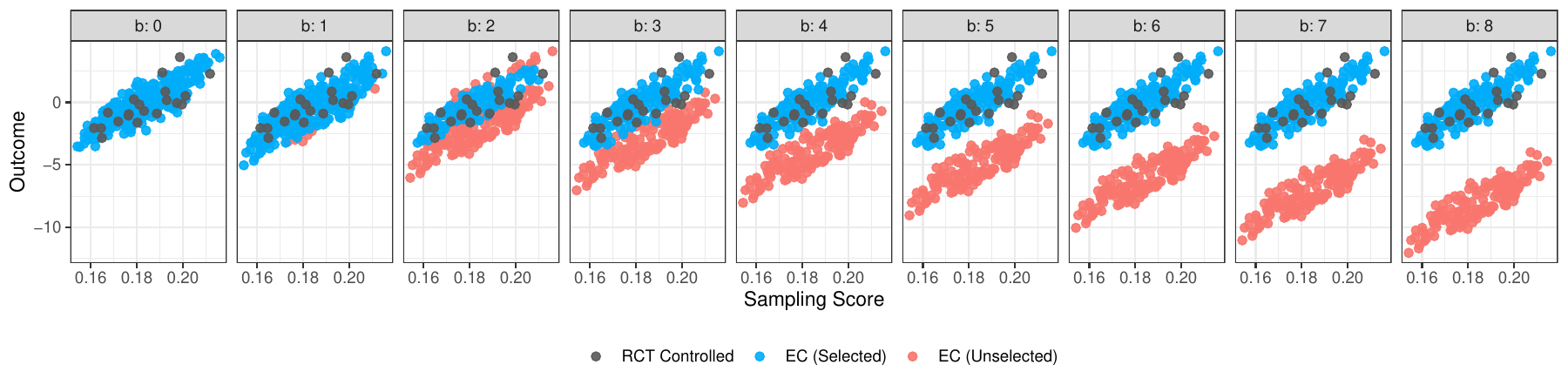}
    \caption{{Selection performance of $\csb$ ($\hg$) when $\ne=300$.}}
    \label{fig:sim_const_ne300_bias}
\end{figure}

\begin{figure}[t]
    \centering
    \includegraphics[width=1\linewidth]{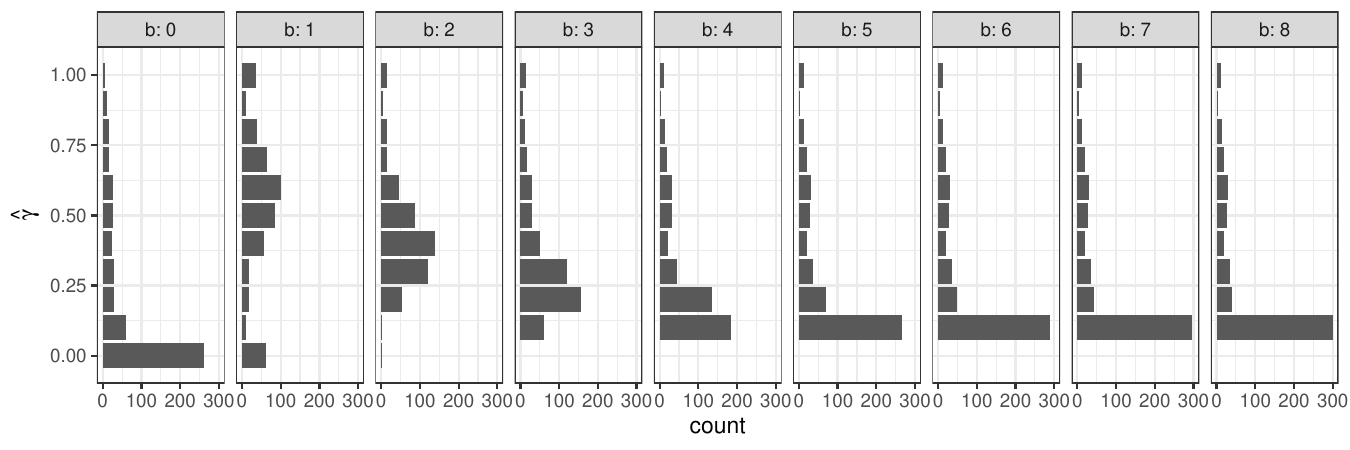}
    \caption{{$\hg$ versus $b$ when $\ne=300$.}}
    \label{fig:sim_const_ne300_adag}
\end{figure}

\begin{figure}[t]
    \centering
    \includegraphics[width=1\linewidth]{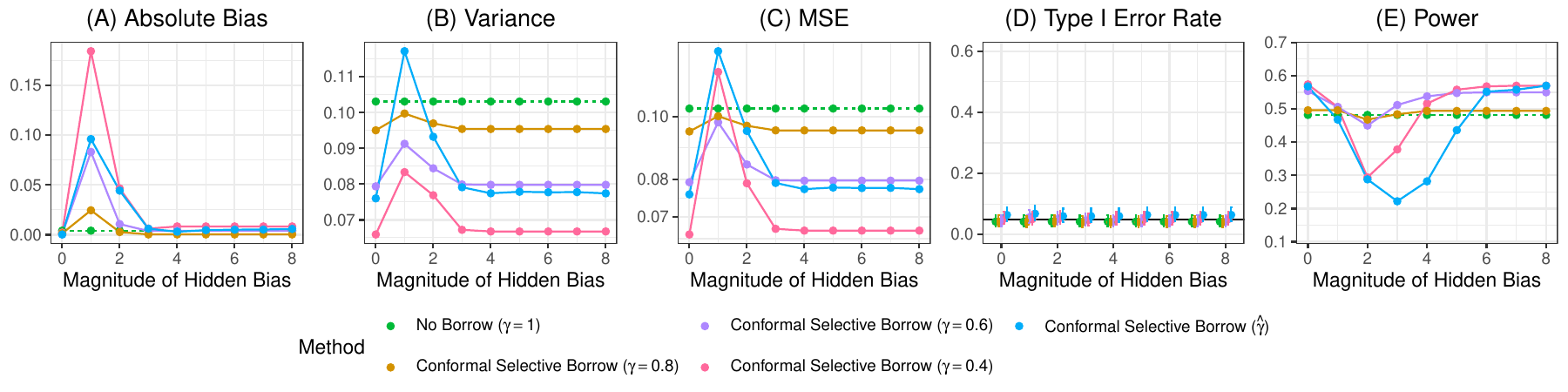}
    \caption{{Simulation results for various selection threshold $\gamma$'s when $\ne=300$.}}
    \label{fig:sim_const_ne300_difg}
\end{figure}

\begin{figure}[t]
    \centering
    \includegraphics[width=1\linewidth]{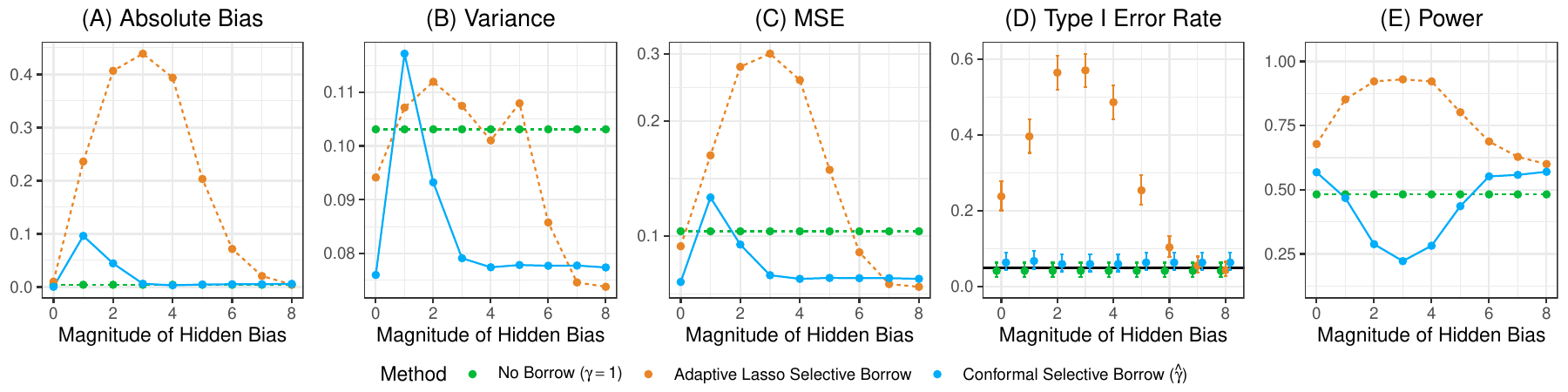}
    \caption{{Comparison of $\csb$ + FRT and $\alsb$ + asymptotic inference when $\ne=300$.}}
\label{fig:sim_const_ne300_alasso}
\end{figure}

Figures \ref{fig:sim_const_ne300_main}, \ref{fig:sim_const_ne300_bias}, \ref{fig:sim_const_ne300_adag}, \ref{fig:sim_const_ne300_difg}, and \ref{fig:sim_const_ne300_alasso} show the simulation results for $\ne=300$. The conclusion is similar to that in the main text.

\subsection{Dependent covariates with $p=5$}

\begin{figure*}[t]
    \centering
    \includegraphics[width=1\linewidth]{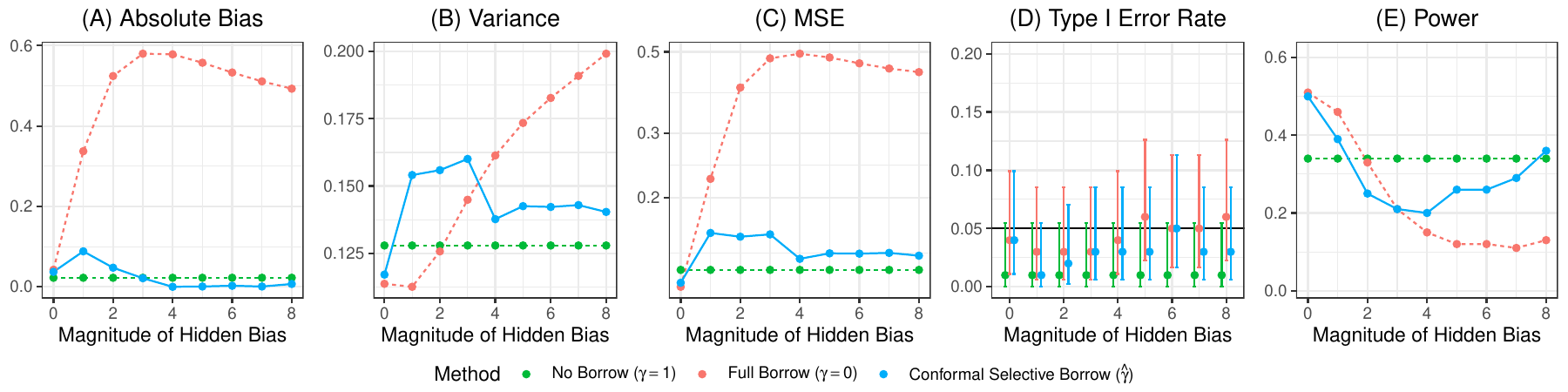}

    \caption{{Simulation results across different hidden bias magnitudes $b$ for dependent covariates with $p=5$.}}
    \label{fig:sim_const_ne50_main2_X}
\end{figure*}

We additionally consider $p = 5$ and $X \sim N(0, \Sigma)$, where $\Sigma$ is a Toeplitz matrix with $\rho = 0.6$ to introduce dependence among the coordinates of $X$. We did not consider larger $p$ since the sample size is small, with only 25 RCT controls. The simulation results (see Figure \ref{fig:sim_const_ne50_main2_X}) show similar conclusions and demonstrate the robustness of our method.

\section{More details about the real data}
\label{sec:addrd}

\begin{figure}[t]
    \centering
    \includegraphics[width=1\linewidth]{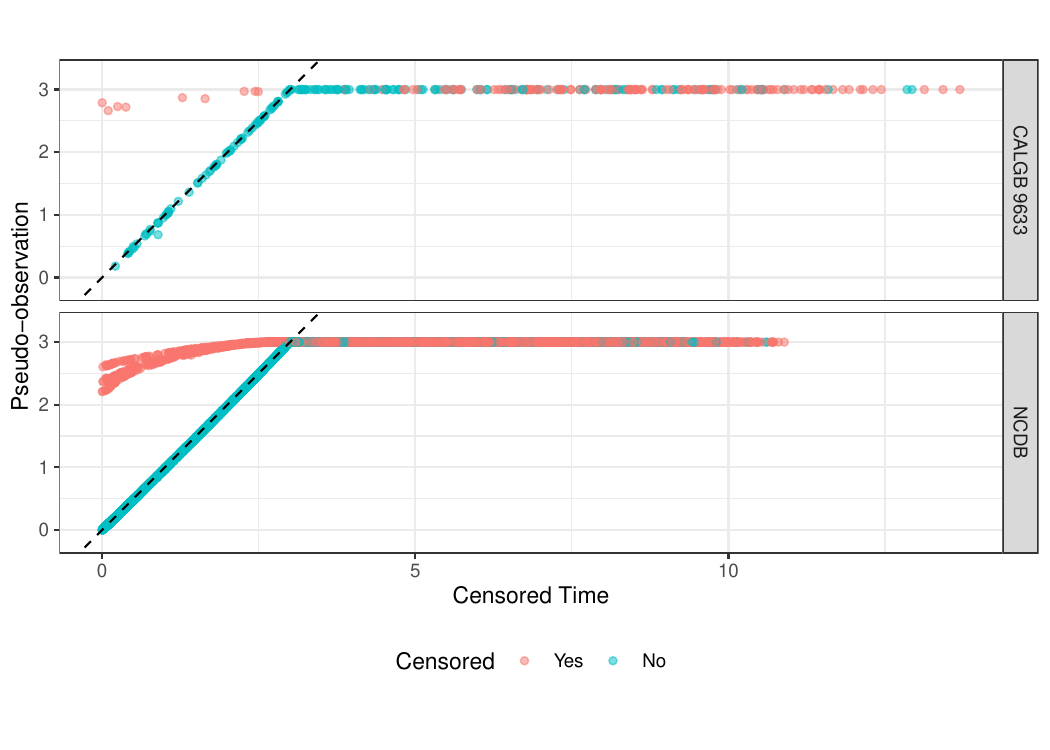}
    \caption{{Pseudo-observation vs. Censored Time for CALGB 9633 and NCDB datasets.}}
    \label{fig:rd_pseudo}
\end{figure}

\textbf{Pseudo-observations.} Figure \ref{fig:rd_pseudo} shows the pseudo-observations versus censored times for CALGB 9633 and NCDB, illustrating that (i) all pseudo-observations are less than or equal to the truncation time of 3 years; (ii) when an event occurs before 3 years, pseudo-observations are generally equal to the event time; and (iii) when censoring occurs before 3 years, pseudo-observations are typically greater than the censored time.

\begin{figure}[t]
    \centering
    \includegraphics[width=\linewidth]{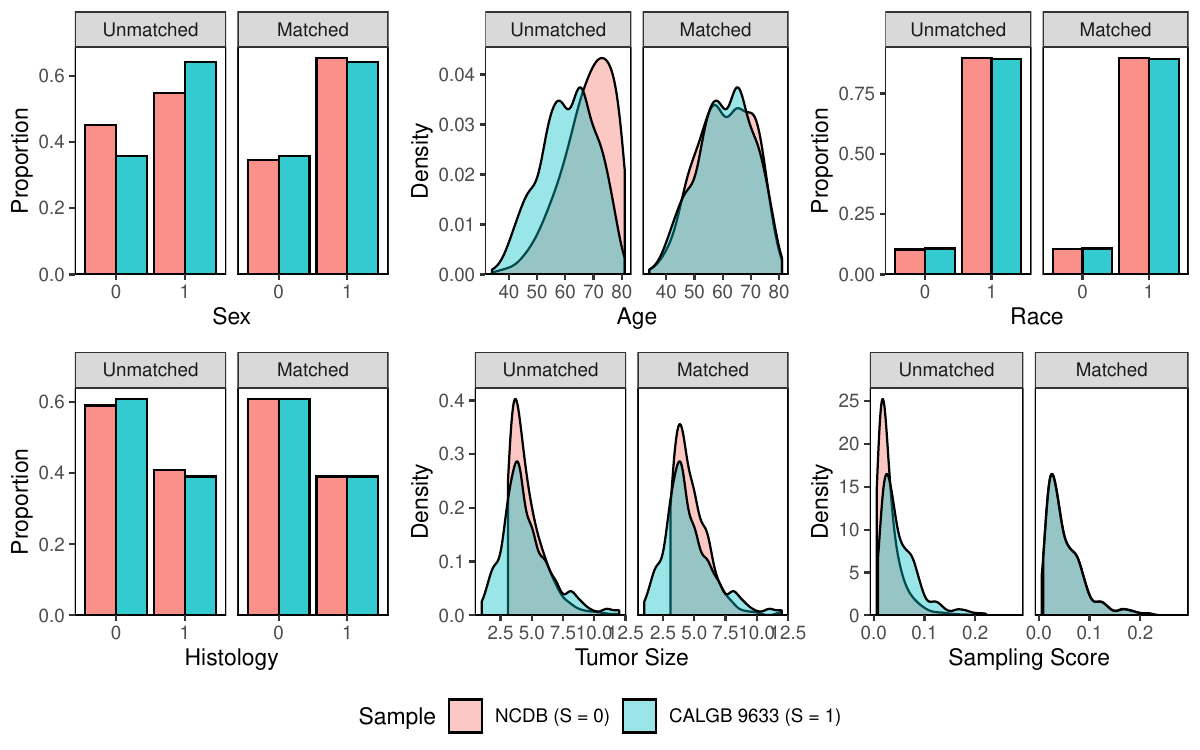}
    \caption{{Distributional balance (unmatched and matched) between CALGB 9633 ($S=1$) and NCDB ($S=0$) for baseline covariates and the estimated sampling score $\hat{\bbP}(S=1|X)$.}}
    \label{fig:rd_cov}
\end{figure}

% matching on X
\textbf{Matching.} We use nearest-neighbor matching to mitigate the covariate imbalance between CALGB 9633 and NCDB. 
Tumor size was imputed for eight missing values in CALGB 9633 using the median of 4. NCDB samples with missing values or covariates outside the CALGB 9633 range were excluded, leaving 10,241 samples.
We perform 1:1 nearest-neighbor matching using \texttt{MatchIt} \citep{ho2011matchit}, treating the sampling indicator $S$ as a ``treatment" and targeting the average treatment effect on the treated (ATT). This preserves all RCT samples and matches 335 NCDB samples.
Distributional balance for the baseline covariates and the estimated sampling score $\hat{\bbP}(S=1|X)$ improves significantly after matching, with a visual comparison in Figure \ref{fig:rd_cov}.
However, certain covariates, such as tumor size, remain imbalanced, which could not be addressed by matching without resorting to methods that would undesirably discard RCT samples. 
This motivates the use of the doubly robust estimator in Sections \ref{sec:pre} and \ref{sec:csb}.
Notably, while a doubly robust estimator alone can address covariate imbalance, matching as a pre-processing step reduces reliance on correct model specification \citep{ho2007matching}.
A summary table of the pre-processed data is in Table \ref{tab:sum}.

\begin{figure}[t]
    \centering
    \includegraphics[width=1\linewidth]{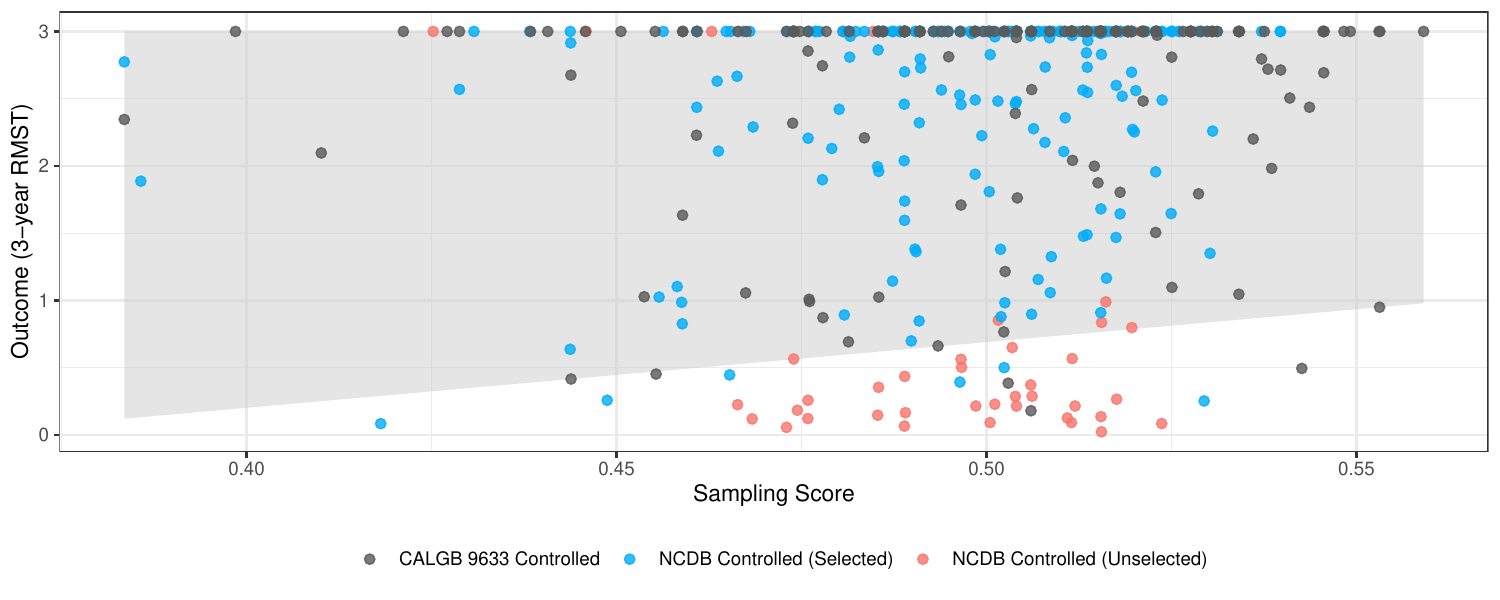}
    \caption{3-year RMST (Outcome) vs. Sampling Score estimated by 5 covariates. The shaded area is constructed using quantile regression on the CALGB 9633 controlled data.}
    \label{fig:rd_ss_sel}
\end{figure}

\textbf{Selection performance.} Figure \ref{fig:rd_ss_sel} shows that, given the observed confounder $X$, $\csb$ tends to select ECs whose outcomes are more similar to randomized controls, reducing hidden bias that cannot be addressed by balancing $X$ alone.

% \textbf{Matching.}  shows that the distributional balance for the five baseline covariates and the estimated sampling score $\hat{\bbP}(S=1|X)$ improve significantly. 

%\textbf{Pre-processing.} Table \ref{tab:sum} shows the summary statistics of the pre-processed data.

\begin{table}
\caption{\label{tab:sum}{Summary statistics of the pre-processed data.}}
\centering
\begin{tabular}[t]{lccc}
\toprule
 & \textbf{C9633 Treated} & \textbf{C9633 Controlled} & \textbf{NCDB Controlled}\\
 & \textbf{($\nt=167$)} & \textbf{($\nc=168$)} & \textbf{($\ne=335$)}\\
 \midrule
\addlinespace[0.3em]
\multicolumn{4}{l}{\textbf{Sex}}\\
\hspace{1em}Male & 109 (65.3\%) & 106 (63.1\%) & 219 (65.4\%)\\
\hspace{1em}Female & 58 (34.7\%) & 62 (36.9\%) & 116 (34.6\%)\\
\addlinespace[0.3em]
\multicolumn{4}{l}{\textbf{Age (years)}}\\
\hspace{1em}Mean (SD) & 60.4 (10.2) & 61.2 (9.28) & 60.8 (9.69)\\
\hspace{1em}Median [Min, Max] & 61.0 [34.0, 78.0] & 62.0 [40.0, 81.0] & 61.0 [34.0, 80.0]\\
\addlinespace[0.3em]
\multicolumn{4}{l}{\textbf{Race}}\\
\hspace{1em}White & 151 (90.4\%) & 148 (88.1\%) & 300 (89.6\%)\\
\hspace{1em}Non-white & 16 (9.6\%) & 20 (11.9\%) & 35 (10.4\%)\\
\addlinespace[0.3em]
\multicolumn{4}{l}{\textbf{Histology}}\\
\hspace{1em}Squamous & 66 (39.5\%) & 65 (38.7\%) & 131 (39.1\%)\\
\hspace{1em}Other & 101 (60.5\%) & 103 (61.3\%) & 204 (60.9\%)\\
\addlinespace[0.3em]
\multicolumn{4}{l}{\textbf{Tumor Size (cm)}}\\
\hspace{1em}Mean (SD) & 4.60 (2.04) & 4.56 (2.05) & 4.77 (1.42)\\
\hspace{1em}Median [Min, Max] & 4.00 [1.00, 12.0] & 4.00 [1.00, 12.0] & 4.50 [3.10, 12.0]\\
\addlinespace[0.3em]
\multicolumn{4}{l}{\textbf{Outcome: 3-year RMST*}}\\
\hspace{1em}Mean (SD) & 2.77 (0.596) & 2.64 (0.720) & 2.43 (0.947)\\
\hspace{1em}Median [Min, Max] & 3.00 [0.383, 3.00] & 3.00 [0.181, 3.00] & 3.00 [0.0242, 3.00]\\
\bottomrule
\multicolumn{4}{l}{\rule{0pt}{1em}*Pseudo-observations transformed from censored survival time.}\\
\end{tabular}
\end{table}

%%%%%%%%%%%%%%%%%%%%%%%%%%%%%%%%%%%%%%%%%%%%%%%%%%%%%%%%%%%%%%%%%%%%%%%%%%%%%%%
%%%%%%%%%%%%%%%%%%%%%%%%%%%%%%%%%%%%%%%%%%%%%%%%%%%%%%%%%%%%%%%%%%%%%%%%%%%%%%%

\end{document}